\documentstyle[11pt,aaspp4]{article}
\def\ie{{\it i.e.,\ }}
\def\eg{{\it e.g.,\ }}
\def\cf{{\it cf.\ }}
\def\etal{{\it et al.\ }}
\def\gtrapprox{\;\lower 0.5ex\hbox{$\buildrel >
    \over \sim\ $}}             
\def\lessapprox{\;\lower 0.5ex\hbox{$\buildrel < \over \sim\ $}}
\def\msol{\ifmmode {\>M_\odot}\else {$M_\odot$}\fi}
\def\pyr{\ifmmode {\>{\rm\ yr}^{-1}}\else {yr$^{-1}$}\fi}
\def\kms{\ifmmode {\>{\rm km\ s}^{-1}}\else {km s$^{-1}$}\fi}
\def\psqcm{\ifmmode {\>{\rm cm}^{-2}}\else {cm$^{-2}$}\fi}
\def\be{\begin{equation}}
\def\ee{\end{equation}}
\def\bea{\begin{eqnarray}}
\def\eea{\end{eqnarray}}
\def\tsigma{\ifmmode {\tilde\sigma}\else {$\tilde\sigma$}\fi}
\def\tomegao{\ifmmode {\tilde\omega_o}\else {$\tilde\omega_o$}\fi}
\def\tsmax{\ifmmode {\tilde\sigma_{i,{\rm max}}}\else
{$\tilde\sigma_{i,{\rm max}}$}\fi}
\def\rthin{\ifmmode {R_{\rm thin}}\else {$R_{\rm thin}$}\fi}
\def\rcirc{\ifmmode {R_{\rm circ}}\else {$R_{\rm circ}$}\fi}
\def\rout{\ifmmode {R_{\rm out}}\else {$R_{\rm out}$}\fi}
\def\nuout{\ifmmode {\nu_{\rm out}}\else {$\nu_{\rm out}$}\fi}
\def\xcirc{\ifmmode {x_{\rm circ}}\else {$x_{\rm circ}$}\fi}
\def\xout{\ifmmode {x_{\rm out}}\else {$x_{\rm out}$}\fi}
\received{Feb. 21, 1996}
\accepted{June 12, 1996}

\slugcomment{To appear in The Astrophysical Journal}

\lefthead{Maloney, Begelman \& Nowak}
\righthead{Warping of Accretion Disks. II.}

\begin{document}

\title{Radiation-Driven Warping. II. Non-Isothermal Disks}

\author{Philip R. Maloney\altaffilmark{1}, Mitchell
C. Begelman\altaffilmark{2,3} and
Michael A. Nowak\altaffilmark{2}}

\altaffiltext{1}{Center for Astrophysics and Space Astronomy,
University of Colorado, Boulder, CO 80309-0389; maloney@shapley.colorado.edu}
\altaffiltext{2}{JILA, University of Colorado and National Institute
of Standards and Technology, Boulder, CO 80309-0440;
mitch@jila.colorado.edu, mnowak@rocinante.colorado.edu} 
\altaffiltext{3}{Also at Department of Astrophysical and Planetary
Sciences, University of Colorado}

\begin{abstract}
Recent work by Pringle and by Maloney, Begelman \& Pringle has shown
that geometrically thin, optically thick, accretion disks are unstable
to warping driven by radiation torque from the central source. This
work was confined to isothermal (\ie surface density $\Sigma\propto
R^{-3/2}$) disks. In this paper we generalize the study of
radiation-driven warping to include general power-law surface density
distributions, $\Sigma\propto R^{-\delta}$. We consider the range
$\delta=3/2$ (the isothermal case) to $\delta=-3/2$, which corresponds
to a radiation-pressure-supported disk; this spans the range of
surface density distributions likely to be found in real astrophysical
disks. In all cases there are an infinite number of zero-crossing
solutions (\ie solutions that cross the equator), which are the
physically relevant modes if the outer boundary of the disk is
required to lie in a specified plane. However, unlike the isothermal
disk, which is the degenerate case, the frequency eigenvalues for
$\delta\neq 3/2$ are all distinct. In all cases the location of the
zero moves outward from the steady-state (pure precession) value with
increasing growth rate; thus there is a critical minimum size for
unstable disks. Modes with zeros at smaller radii are damped. The
critical radius and the steady-state precession rate depend only
weakly on $\delta$.

An additional analytic solution has been found for $\delta=1$.  The
case $\delta=1$ divides the solutions into two qualitatively different
regimes. For $\delta \ge 1$, the fastest-growing modes have maximum
warp amplitude, $\beta_{\rm max}$, close to the disk outer edge, and
the ratio of $\beta_{\rm max}$ to the warp amplitude at the disk inner
edge, $\beta_o$, is $\gg1$. For $\delta < 1$, $\beta_{\rm
max}/\beta_o\simeq 1$, and the warp maximum steadily approaches the
origin as $\delta$ decreases. This implies that nonlinear effects {\it
must} be important if the warp extends to the disk inner edge for
$\delta \ge 1$, but for $\delta < 1$ nonlinearity will be important
only if the warp amplitude is large at the origin. Because of this
qualitative difference in the shapes of the warps, the effects of
shadowing of the central source by the warp will also be very
different in the two regimes of $\delta.$ This has important
implications for radiation-driven warping in X-ray binaries, for which
the value of $\delta$ characterizing the disk is likely to be less
than unity.

In real accretion disks the outer boundary condition is likely to be
different from the zero-crossing condition that we have assumed. In
accretion disks around massive black holes in active galactic nuclei,
the disk will probably become optically thin before the outer disk
boundary is reached, while in X-ray binaries, there will be an outer
disk region (outside the circularization radius) in which the inflow
velocity is zero but angular momentum is still transported. We show
that in both these cases the solutions are similar to the
zero-crossing eigenfunctions.
\end{abstract}


\section{Introduction}
Evidence for warped, precessing accretion disks in astrophysical
systems ranging from X-ray binaries to active galactic nuclei has
steadily accumulated over the last two decades (see Maloney \&
Begelman 1997a, and references therein). The origin and maintainence
of such warped disks has until recently stood as an unsolved
theoretical problem. While it is possible, for example, to generate
non-planar modes with $m=1$ symmetry in thin, relativistic disks
(\cite{k90}; \cite{kh91}), these modes only exist at small radii
($R\lessapprox 10$ Schwarzschild radii), since they rely on trapping
of the modes in the non-Newtonian region of the potential. However, an
important clue was provided by \cite{pet77}, who pointed out that in
an optically thick disk with a central source of luminosity, the
pressure resulting from re-radiation of the intercepted flux will
produce a net torque if the disk is warped.  Almost twenty years were
to pass before it was recognized that radiation pressure torque
actually leads to a warping instability.  \cite{pri96} (P96) showed
that, for the special case in which the disk surface density
$\Sigma\propto R^{-3/2}$ (corresponding to an isothermal disk in the
usual $\alpha-$disk formalism, with disk viscosity $\nu\propto
R^{3/2}$), even an initially planar disk is unstable to warping by
this mechanism. Pringle solved the linearized twisted disk equations
in this case using a WKB approximation.  Maloney, Begelman \& Pringle
(1996, Paper I, hereafter MBP) found exact solutions to the linearized
twist equations, and demonstrated the importance of the outer boundary
condition for determining the growth rates of the unstable modes.

These previous works all specialized to the case of an isothermal disk,
which simplifies the twist equations. While this may be a reasonable
approximation for some astrophysical disks (\eg the masing molecular
disk in NGC 4258; see MBP), there are many other systems, such as
accretion disks in X-ray binary systems, where this is likely to be a
poor assumption. In this paper we extend the work of MBP by
considering disks with power-law surface density profiles,
$\Sigma\propto R^{-\delta}$. We consider the range $-3/2\le\delta\le
3/2$: the lower limit corresponds to a radiation-pressure-supported
disk (\eg \cite{fkr92}), while the upper limit is the
isothermal value (MBP). Within the limitations of assuming a constant
power-law for the surface density, this spans the probable range of
surface density laws relevant to real astrophysical accretion
disks. For example, the standard Shakura-Sunyaev gas
pressure-supported disk is characterized by $\delta=0.75$
(\cite{ss73}).

In \S 2 we discuss the twist equation, including the effect of
radiation torque, and cast it into a more convenient form. We solve
the equation numerically in \S 3 and discuss both the time-dependent and 
steady-state solutions. As in the isothermal case, the outer boundary
condition is crucial for determining the stability of the disk and the
growth rates of the unstable modes. In \S 4 we discuss the important
issue of the appropriate outer boundary condition for accretion disks
around stellar-mass objects and AGN. Finally, in \S 5 we discuss the
implications of the results and their application to real accretion
disks.

\section{The Disk Evolution Equation}
As in P96 and MBP, we adopt a Cartesian coordinate system with the
$Z-$axis aligned with the normal to the unwarped disk, and define
$\beta$ to be the local angle of tilt of the disk axis with respect to
the $Z-$axis, while $\gamma-\pi/2$ is the angle between the descending
line of nodes and the $X-$axis. The equation governing the evolution
of the local tilt vector, ${\bf l}(R,t)=(\sin\beta\cos\gamma,
\sin\beta\sin\gamma, \cos\beta)$, including the radiation torque term,
is then (P96)
\be
{\partial{\bf l}\over\partial t}+\left[V_R -{\nu_1\Omega^\prime\over
\Omega}-{1\over 2}\nu_2{\left(\Sigma R^3\Omega\right)^\prime\over 
\Sigma R^3\Omega}\right]{\partial{\bf l}\over \partial
R}={\partial\over\partial R}\left({1\over 2}\nu_2{\partial{\bf
l}\over\partial R}\right)+{1\over 2}\nu_2\left|{\partial{\bf
l}\over\partial R}\right|^2{\bf l}+{{\cal G}\over \Sigma R^2\Omega}
\ee
where $\nu_1$ and $\nu_2$ are the disk viscosity in the azimuthal and
vertical directions (with a ratio $\eta\equiv \nu_2/\nu_1$ that is
assumed constant but not necessarily unity), primes denote derivatives
with respect to $R$, $\Sigma$ is the disk surface density, $\Omega$
the Keplerian angular velocity, $V_R$ the radial inflow speed, and the
radiation torque term 
\be 
{\cal G}\equiv {1\over 2\pi R}{d{\bf G}\over dR} ,
\ee
where $d{\bf G}$ is the torque exerted on a ring of width $dR$ and
radius $R$ and is given by equation (2.18) of P96.

As in MBP, we assume the disk viscosity can be written $\nu_1(R)=\nu_o 
(R/R_o)^\delta$, where $R_o$ is an arbitrary fiducial radius, but we
now allow the radial power-law to be arbitrary, rather than
specializing to the case $\delta=3/2$. In a
steady-state disk far from the boundaries, $\Sigma=\dot M/3\pi\nu_1$,
so this radial dependence of the viscosity translates directly into a
power-law surface density, $\Sigma\propto R^{-\delta}$. Furthermore, in
a steady disk the radial inflow velocity
$V_R=\nu_1\Omega^\prime/\Omega$, and so the first two terms inside the
brackets on the LHS of equation (1) cancel, while the third term
becomes
\be
{\left(\Sigma R^3 \Omega\right)^\prime\over \Sigma R^3 \Omega}
=\left({3\over 2}-\delta\right){1\over R}\;.
\ee
We then linearize to obtain the more general version of equation (1)
of MBP for $W\equiv\beta e^{i\gamma}$,
\be
{\partial W\over \partial t}=\left({3\over 4}{\nu_2\over
R}-i\Gamma\right) {\partial W\over \partial R} 
+\left({1\over 2}\nu_2 {\partial^2 W\over
\partial R^2}\right) ,
\ee 
where the radiation torque term is
\be
\Gamma={L\over 12\pi\Sigma R^2\Omega c}
\ee
with $L$ the luminosity of the central source. Assuming an
accretion-fueled source with radiative efficiency $\epsilon\equiv
L/\dot M c^2$, this term can be written as 
\be
\Gamma={\epsilon\over 2\sqrt 2 R_s^{1/2}}{\nu_o\over
R_o^{\delta}}R^{\delta-1/2}\equiv\Gamma_o R^{\delta-1/2}
\ee
where $R_s$ is the Schwarzschild radius (cf. equation [3] of
MBP). Transforming to the new radius variable 
$x=a R^{1/2}$, using equation (6) for $\Gamma$, and Fourier
transforming with respect to time, the twist evolution equation
becomes
\be
{\partial W\over \partial t}=i\sigma W={\eta\nu_o\over 8
R_o^\delta}a^{4-2\delta}x^{2\delta-3}\left[x{\partial^2 W\over \partial x^2}+
\left(2-{i\sqrt 2\epsilon\over\eta a R_s^{1/2}}x\right){\partial
W\over\partial x}\right]\ee
where in general $\sigma$ is complex.
We set $a=\sqrt 2\epsilon/R_s^{1/2}\eta$ (note that this differs from
the definition of radius variable $z$ in MBP only in making $x$ real,
rather than pure imaginary). Since $R_o$ is arbitrary, we now define
$R_o$ so that $x(R_o)=1$; $\nu_o$ is thus the value of $\nu_1$ at
$x=1$. Therefore
\be
R_o=a^{-2}= {R_s \eta^2\over 2 \epsilon^2}
\ee
and the coefficient of the $x^{2\delta-3}$ term on the righthand
side of equation (7) becomes $\nu_o \epsilon^4/2 R_s^2\eta^3$. Finally,
we nondimensionalize the eigenfrequency $\sigma$ by defining
\be
\tsigma\equiv {2\eta^3 R_s^2\over \epsilon^4\nu_o}\sigma
\ee
to obtain the final form of the twist equation:
\be x{\partial^2 W\over \partial x^2}+
(2-ix){\partial W\over\partial x}-i\tsigma x^{3-2\delta} W=0\;.
\ee
For $\delta=3/2$, equation (10) reduces to Kummer's equation, as
discussed in MBP. (Note that the coefficient of $W$ in equation [9] of
MBP, $2\sigma/\Gamma_o$, is identical to $\tsigma$ as defined here.)
Equation (10) can thus be regarded as a modified Kummer's
equation. Unfortunately, this equation is in general analytically
intractable, so that numerical solution is necessary. We discuss the
asymptotic behavior of the solutions in Appendix A. For the physically
relevant solutions (those which exhibit zero-crossings), there are in
all cases an infinite number of zero-crossing eigenfunctions. As we
will see below, the special case $\delta=3/2$ is degenerate: the real
parts of the eigenvalues are all identical, with ${\rm
Re}(\tsigma)=1$. For all the other values of $\delta$ that we consider the
eigenvalues are distinct, and each eigenfunction has at most one zero.
In addition, the real part of $\tsigma$, $\tilde\sigma_r$, must be
greater than zero, which means that the direction of precession of the
warp must be the same as the direction of disk rotation (\ie prograde:
see Appendix B). This is a simple consequence of the overall shape of
the warp: it is easy to show that for a disk in which the gradient in
the tilt $\beta' > 0$, the direction of precession due to the radiation
torque is retrograde, while if $\beta' < 0$ the precession will be
prograde. Since the solutions are constrained to return to the plane
at the outer boundary, $\beta'$ is always negative at large radius,
and this dominates the direction of the induced precession.

We have found one additional analytic solution to equation (10), for
$\delta=1$ (Appendix C). We discuss this and the numerical solutions
to equation (10) for both steady-state and unstable modes in the
following section.

\section{Solutions of the Twist Equation}
Real astrophysical accretion disks will generally be unwarped beyond
some maximum radius $R_{\rm max}$, either because the disk becomes
optically thin to the incident or re-emitted radiation (see the
discussion in MBP), so that the disk must eventually return to the
initial plane, or because this is forced in some other manner by the
physical outer boundary condition (\eg if the accretion disk is fed by
material from a companion star; see \S 4.2). We therefore consider
solutions of the twist equation for which the disk returns to the
original plane $Z=0$ at some radius. This outer boundary condition is
not rigorously correct, as we discuss in \S 4, but it is in general an
excellent approximation to the true outer boundary condition, and the
zero-crossing eigenfunctions are very useful for understanding the
behavior of the warping instability.

Equation (10) is a second-order differential equation, and therefore
requires two boundary conditions to specify the solution. In principle
it could be solved as a two-point boundary value problem, except that
the location of the outer boundary is not necessarily known {\it a
priori}. It is computationally most convenient to solve it as an
initial-value problem, making an initial guess for the value of
\tsigma\ and then iterating to find the solution that goes to
zero. We separate equation (10) into a coupled pair of equations for
the real and imaginary parts of $W$, and integrate outward from the
origin using a fifth-order Runge-Kutta scheme (\cite{nr92}).

We require solutions that are regular at the origin. As $x\rightarrow
0$, the leading terms of equation (10) are
\be
x W^{\prime\prime} + 2 W^\prime = i x^{3-2\delta}\tsigma W
\ee
where primes now denote derivatives with respect to $x$. Adopting the
boundary condition $W(0)=1$, the leading behavior of (11) as
$x\rightarrow 0$ gives
\be
W^\prime \rightarrow {i\tsigma\over {5-2\delta}}
x^{3-2\delta}\quad\quad{\rm as}\quad x\rightarrow 0.
\ee
Furthermore, we require that zero torque be exerted on the disk
at the origin, which requires that $x^2 W^\prime\rightarrow 0$ as
$x\rightarrow 0$. From equation (12), it is evident that this second
boundary condition is satisfied for any $\delta < 5/2$. 

Numerical solutions to the twist equation are presented
below. However, we first note that there are simple scaling relations
that are generic features of the instability:
\begin{itemize}
\item[(1)] The radiation torque per unit area is $\Gamma\sim
(L/4\pi R^2c)\times R$, while the angular momentum density is $l\sim
\Omega R^2\Sigma\sim \Omega R^2\dot M/3\pi\nu_1$, where $\Omega$ is the
angular velocity at radius $R$ in the disk, $\dot M$ is the mass
accretion rate, and $\nu_1$ is the usual kinematic viscosity. The radiation
torque timescale is thus given by
\begin{equation}
t_{\rm rad}\sim {\Omega R^3\over c\nu_1}{\dot M c^2\over L}
={\Omega R^3\over c\nu_1}\;\epsilon^{-1}
\end{equation}
which, for an accretion-fueled source, depends only on the accretion
efficiency $\epsilon=L/\dot M c^2$ and not on the luminosity $L$ and
the mass accretion rate individually (MBP; Maloney \& Begelman 1997a).
\item[(2)] The viscous timescale is $t_{\rm visc}\sim R/V_R=2R^2/3\nu_1$, so
the ratio of viscous to radiation torque timescales is given by
\begin{equation}
{t_{\rm visc}\over t_{\rm rad}}\sim {\epsilon c\over\Omega R}\sim
\epsilon\left({R\over R_g}\right )^{1/2}\;;\quad\quad
R_g={GM\over c^2}
\end{equation}
which is independent of the form of the viscosity law and depends only
on $\epsilon$ and the radius normalized to the 
gravitational radius. Thus the radiation torque always wins out at
large radii, \ie for $R > \epsilon^{-2} R_g$, but viscosity always
dominates near the center. This is why the disk will be {\it flat}
(but in general will have non-zero tilt) at small radii. Note that
this equation also implies that ${t_{\rm visc}/t_{\rm rad}}\sim 1$ at
$x=1$. 
\item[(3)] As is apparent from equations (7) -- (9), altering the values of
$\eta$ and $\epsilon$ while keeping the ratio $\eta/\epsilon$ fixed
will affect the growth rates, but not the {\it shapes} of the
solutions, \ie the value of $x$ is unaffected.
\end{itemize}

\subsection{Steady-State Solutions}
We first discuss the steady-state solutions to the twist equations,
for which $\tsigma_i\equiv{\rm Im}(\tsigma)=0$.

Figure 1 shows the location of the zero, $x_o$, for the first ten
zero-crossing eigenfunctions, as a function of surface density index
$\delta$. As noted earlier, for $\delta\neq 3/2$ the eigenfunction
corresponding to each eigenvalue has only a single zero (Appendix
B). We use ``order'' to specify how many eigenfunctions have their
zero at values of $x$ smaller than or equal to the eigenfunction in
question; thus the eigenfunction with the smallest value of $x_o$ is
the first-order eigenfunction, that with the next smallest is the
second-order eigenfunction, and so forth. The behavior of the
higher-order eigenfunctions is remarkably complex.

In Figure 2 we plot the normalized real eigenvalues, $\tsigma_r$,
for the first ten order eigenfunctions. The first order eigenfunction
has the largest eigenvalue for all values of $\delta$. The degeneracy
of the case $\delta=3/2$ discussed in MBP is immediately apparent,
with $\tsigma_r=1$ for all the eigenfunctions. For $\delta\neq 3/2$,
the degeneracy is lifted, with increasing separation of the
eigenvalues as $\delta$ decreases. Closer examination shows that the
behavior becomes quite complex for the higher-order zeros with
decreasing $\delta$, as merging of eigenvalues occurs.

The location of the first-order zero as a function of $\delta$ has an
important physical significance. In real astrophysical disks,
shadowing effects probably make the higher-order eigenfunctions
unimportant. Just as for the isothermal case, for the growing modes
the location of the zero moves outward from the steady-state value as
the growth rate increases (see below); all the modes with zeros at
smaller radius are damped. Thus the location of the first zero marks
the critical boundary for disk stability. Accretion disks larger than
\be R_{\rm cr}={1\over 2}\left({\eta\over\epsilon}\right)^2 x_{\rm
cr}^2 R_s \ee are unstable to warping by radiation pressure, where the
critical value $x_{\rm cr}$ is equal to $2\pi$ for $\delta=3/2$ and
$x_{\rm cr}\simeq 4.891\pi$ for $\delta=-3/2$. Figure 1 shows that
$x_{\rm cr}$ increases smoothly as $\delta$ decreases. Disks with
outer boundaries smaller than $R_{\rm cr}$ are stable against
radiation-driven warping. From equation (15), this critical radius
scales as $\epsilon^{-2}$, so that accretion disks in low-efficiency
systems will not be unstable unless the disks are extremely
large. (Furthermore, the precession and growth timescales for the
instabilty will be extremely long if $\epsilon$ is very small; \cf
the discussion of equations [30] and [31] in \S 5.) 

In Figure 3 we plot the tilt $\beta$ as a function of radial variable
$x$ for the first-order steady-state solutions, for several values of
$\delta$. For clarity they have been plotted only to the zero of
the mode. In all cases the maximum tilt is at the origin (note that
$\beta_{\rm max}$ is arbitrary and has been taken to be unity); as
$\delta$ decreases from 3/2 the zero moves to larger $x$ and an
increasing fraction of the disk has $\beta\simeq\beta_{\rm max}$. 

The steady-state modes have shapes that are time-independent in a frame
rotating with angular velocity $\sigma_r$; physically, these are
purely precessing modes. The frequency $\sigma_r$ is related to the
dimensionless frequency $\tsigma_r$ by equation (9). As can be seen
from Figure 2, $\tsigma_r$ for the first-order zero drops by six
orders of magnitude from $\delta=3/2$ to $\delta=-3/2$. However, to
put the results in physical terms, note that from the definition of
$x$, we can rewrite equation (9) as
\be
\sigma={\nu_o\eta\over 8 R_o^2}\tsigma\;.
\ee
Defining the viscous timescale as before as 
\be
t_{\rm visc}(R)={2 R^2\over 3\nu_1}\;,
\ee
we can then express $\sigma$ in terms of the viscous timescale at the
critical radius,
\be
\sigma={\eta x_{\rm cr}^{4-2\delta}\over 12 t_{\rm visc}(R_{\rm
cr})}\tsigma
\ee
where $x_{\rm cr}$ is the corresponding value of $x$ (\cf equation
[15]). For $\delta=-3/2$, $x_{\rm cr}^{4-2\delta}\tsigma_r=176.8$,
while for $\delta=+3/2$ this quantity equals $2\pi$, so that, in terms
of the viscous timescale at $R_{\rm cr}$, the variation in $\sigma_r$
is only a factor of $\approx 28$ over the entire range of
$\delta$. Alternatively, $\sigma$ can be expressed in terms of the
viscous timescale at the outer edge of the disk (\ie the location of
the zero), with $x_{\rm cr}$ and
$R_{\rm cr}$ replaced by $x_o$ and $R(x_o)$. (In fact,
the variation in $\sigma_r$ is smaller than this, because for a
fixed value of the viscosity, the increase in $R_{\rm cr}$ with
decreasing $\delta$ causes $t_{\rm visc}(R_{\rm cr})$ to increase,
offsetting the variation in $x_{\rm cr}^{4-2\delta}\tsigma_r$; see \S
5.)

\subsection{Unstable Solutions}
We now consider solutions that evolve with time, \ie $\tsigma_i\neq
0$. As for the isothermal disks analyzed in MBP, the only modes that
have zeros at radii smaller than $x_{\rm cr}$, as discussed above,
are damped, with $\tsigma_i > 0$, and we do not consider them
further. 

For the case of an isothermal disk considered by MBP, the zeros move
steadily outward with increasing growth rate (\ie increasing
$-\tsigma_i$), and there is no limit to the growth rate. For the
non-degenerate cases considered here, the zeros also move outward with
increasing $-\tsigma_i$, as expected from the scalings discussed in \S
3. However, since the degeneracy of the eigenvalues has been lifted,
there is now a maximum growth rate for a given order mode. In
particular, in Figure 4, we plot the location of the first-order zero,
$x_o$, (as discussed above, this is probably the only physically
relevant eigenfunction) as a function of $-\tsigma_i$, for
$\delta=-3/2$ to $1.45$, in steps of $0.05$ in $\delta$. The
systematic behavior of the location of the zero is quite striking, and
there is a clear transition across $\delta=1$. The maximum possible
(normalized) growth rate $-\tsigma_i$ for the first-order zero
steadily decreases as $\delta$ decreases. But the location of the zero
at this maximum growth rate, $-\tsmax$, moves to larger $x$ as
$\delta$ decreases from $1.45$ to $1.00$, even though $-\tsmax$
systematically declines with decreasing $\delta$.

The case $\delta=1$ is special. There is an analytic solution in this
case, as discussed in Appendix C. The maximum growth rate is
$-\tsmax=0.25$; at this value of $\tsigma_i$, the zero has moved to
$x=\infty$. (The curve plotted in Figure 4 ends at $-\tsmax=0.249$.)
This value of $\delta$ is also special in that all of the eigenvalues
merge as $\tsigma_i\rightarrow\tsmax$, so that the zeros of all orders
move to infinity at $-\tsmax=0.25$. Figure 5a plots the location of
the first three order zeros for $\delta=1$ as a function of $-\tsigma_i$,
while Figure 5b shows the real parts of the corresponding eigenvalues.

For $\delta < 1$, the maximum $x-$value for the first-order zero
decreases rapidly as $\delta$ decreases, reaching a minimum at
$\delta\simeq 0.15$, and then slowly increasing as $\delta$ approaches
$-3/2$ (see Figure 4). The ratio $x_o(\tsmax)/x_o(\tsigma_i=0)$
decreases as $\delta$ decreases, so that the range of instability (out
to the first zero) decreases with $\delta$ in this range.

In Figure 6 we plot the real part of the eigenvalue, $\tsigma_r$,
versus the imaginary part for the growing modes, for the same values
of $\delta$ as in Figure 4. In all cases $\tsigma_r$ shows a
characteristic decrease as \tsmax\ is approached. As in Figure 4, the
curve for $\delta=1$ is plotted only to $-\tsigma_i=0.249$;
$\tsigma_r\rightarrow 0$ as $-\tsigma_i\rightarrow 0.25$. The
magnitude of the drop in $\tsigma_r$ declines as $\delta$ approaches
$3/2$; for the isothermal case, $\tsigma_r=1$, independent of
$\tsigma_i$ (MBP). 

There is a dramatic change in the behavior of the unstable solutions
across $\delta=1$. In Figure 7 we have plotted the shape of the
eigenfunctions, \ie $\beta$ as a function of $x$, for several values
of $\delta > 1$, for the fastest-growing modes, with
$\tsigma_i=\tsmax$. The eigenfunctions have been normalized so that
the maximum value of $\beta$ is unity; the normalization factor (\ie
$1/\beta_{\rm max}$) is very small, of order $10^{-6}$. The behavior
of the eigenfunctions is very similar to the fast-growing modes of the
isothermal disks discussed in MBP: the warp has its maximum close to
the outer boundary of the disk (assumed to be coincident with the
zero), and the amplitude of the warp at this maximum is very large
compared to that at $x=0$ (approximately $10^6$ for the modes shown in
Figure 7). The shape of the fastest-growing modes for $\delta=1$ are
qualitatively similar to the $\delta > 1$ modes; the only difference
is that $x_o$ approaches $\infty$ (rather than a finite value) as
$\tsigma$ approaches \tsmax.

In Figure 8 we similarly plot the normalized eigenfunctions for the
fastest-growing modes, for several values of $\delta < 1$. The change
in the nature of the eigenfunctions is marked: even for $\delta=0.95$,
the amplitude of $\beta$ at the maximum is only $\sim 2.5$ times the
value of $\beta$ at $x=0$, $\beta_o$, and as $\delta$ decreases
$\beta_{\rm max}/\beta_o$ declines toward unity. For $\delta
\lessapprox 0.5$, the eigenfunctions converge to an essentially
constant shape, with a plateau of constant $\beta$ extending from the
origin to $x\sim 5$, followed by a decrease to the zero. This change
in the behavior of the eigenfunctions across $\delta=1$ has important
implications, as we discuss below. Note also that this applies only to
the unstable modes; the steady-state solutions do not exhibit any
change in behavior across $\delta=1$, as is apparent from Figure 3.

\section{Outer Boundary Conditions}
In section 3, we solved equation (10) with the requirement that the
solution go to zero at the disk outer boundary, which is at some
finite radius. While this produces a well-defined set of solutions
whose behavior is very useful for understanding the behavior of the
instability, in real astrophysical disks the true outer boundary
condition is likely to be different. Furthermore, the correct outer
boundary condition for accretion disks around stellar-mass compact
objects (neutron stars and black holes) in X-ray binaries will differ
from that appropriate for accretion disks around massive black holes
in active galactic nuclei. In this section we discuss the choice of
outer boundary condition and how the solutions in these cases differ
from the zero-crossing solutions discussed above.

\subsection{Active Galactic Nuclei}
It is expected that accretion disks will generally satisfy the
requirement of optical thickness, and will therefore be subject to the
radiation-warping instability (see \S 5 and Appendix D). However,
accretion disks surrounding massive black holes in active galactic
nuclei are likely to become optically thin (to the re-emitted
radiation) before the physical boundary of the accretion disk is
reached. (This is true for $\delta > 0$, so that the disk surface
density decreases with increasing radius; this is undoubtedly the case
at the relevant radii of AGN accretion disks.) Once the disk becomes
optically thin, the instability ceases to operate, as the re-emitted
radiation no longer exerts a torque on the disk (\cf MBP). Therefore,
equation (10) does not describe the behavior of the disk beyond the
optically thin radius \rthin. However, viscosity continues to operate,
so that twist angular momentum generated by the instability can be
transported beyond \rthin. We therefore have to solve a pair of
equations, with the solutions matched at \rthin: interior to \rthin,
the equation governing the dynamics of the disk is equation (10),
while exterior to \rthin\ the disk is described by equation (10) minus
the torque term $-ixW'$. Both $W$ and $W'$ must be continuous at
\rthin; the outer disk solution must also obey $W\rightarrow 0$ as
$R\rightarrow \infty$. In a real accretion disk, the radiation torque
will presumably decline smoothly to zero rather than switching off
abruptly, but as long as this occurs in a radial thickness $\Delta
R\ll R$ we expect this to have little effect on the solutions.

For $\delta=3/2$ and $\delta=1$ the outer disk equation can be solved
analytically; the solutions are modified Bessel functions ($K_1$ and
$K_{1/2}$, respectively). In general it must be solved numerically. For a
specified value of \rthin\ we solve both the inner and outer disk
equations and iterate to find the solution that satisfies the
matching condition at \rthin. (In practical terms, the constraint
that must be satisfied is that $\beta'$ be continuous at \rthin,
since $\beta$ can be scaled arbitrarily and $\gamma$ does not enter
into equation (10), only $\gamma'$ and $\gamma''$; the value of
$\gamma$ can always be matched by an arbitrary rotation of the outer
disk solution.) Since the parameter space to be considered is very
large, we consider only a few representative solutions for
$\delta=3/2$, $1.25$, and $0.75$; the behavior of solutions for
$\delta < 0.75$ is very similar to that of the $\delta=0.75$ solutions.

We first consider the modifications to the steady-state, first-order
solutions shown in Figure 3. We take \rthin\ to be the zero of these
modes, and then find the solutions that satisfy the boundary
conditions at \rthin. In Figure 9 we plot the real versus the
imaginary parts of the eigenvalues for these solutions. Not
surprisingly, there is now a continuum of solutions rather than a
single mode, as it is easier to satisfy the boundary condition
$\beta\rightarrow 0$ as $x\rightarrow\infty$ rather than at a finite
value of $x$. The solid lines correspond to growing modes ($\tsigma_i
<0$) while the dotted portions of the curves are damped. The shapes of
the solution curves in the $(\tsigma_r,\tsigma_i)$-plane are very
similar for all values of $\delta$, with a pair $\tsigma_r$-values for
a given value of $\tsigma_i$. In all cases there is a well-defined
maximum (and minimum) growth rate at which the two values of
$\tsigma_r$ merge; the precession rate $\tsigma_r$ corresponding to
this fastest-growing mode differs from that of the zero-crossing
solution by $\sim 2\%$, $8.5\%$, and $13\%$ for $\delta=1.5$, $1.25$,
and $0.75$, respectively.

In Figure 10 we plot the tilt $\beta$, normalized to a maximum of
unity as before, for the steady-state zero-crossing modes (dotted
lines) and the corresponding fastest-growing modes from Figure 9, with
$\rthin=x_o$ (solid lines). The behavior of the ``AGN'' modes is what
one would expect: since the growth rates are non-zero, the maximum of
the tilt has moved outward from the origin, and the solutions decline
smoothly to zero, with $W'\rightarrow 0$ as $x\rightarrow\infty$.
Since the differences in shape between the steady-state mode and
growing modes are minor for $\delta \lessapprox 0.75$, as discussed in
\S 3, the differences between the zero-crossing modes and the
optically-thin-boundary modes are minimal for the $\delta=0.75$
case. We also note that beyond \rthin\ $\gamma'$ becomes negative (\ie
the line of nodes describes a retrograde spiral), since in the absence
of radiation torque viscosity simply causes the line of nodes to
unwind.

We now consider modifications to the rapidly growing modes. Figure 11
plots the precession rate versus the growth rate for rapidly growing
modes for the three values of $\delta$. For $\delta=1.25$ and $0.75$
the zero-crossing modes used to set the value of \rthin\ have growth
rates close to the maximum for first-order modes. For $\delta=1.50$
there is no maximum growth rate (due to the degeneracy) and so the
solution shown is simply a rapidly growing mode, with
$\tsigma_i=-5$. The behavior of the continuum of modes that satisfies
the boundary conditions is very similar to those shown in Figure 9,
except that there are no damped modes in this case. Comparison with
Figure 9 also shows that for $\delta\neq 1.50$, the width of the curve
in the $(\tsigma_r,\tsigma_i)$-plane is significantly smaller for
these fast-growing modes. The precession frequencies of the
maximally-growing modes are even closer to those of the zero-crossing
modes than for those shown in Figure 9, differing by $\sim 0.05\%$,
$1\%$, and $8\%$ in order of decreasing $\delta$.

In Figure 12 we plot $\beta$ as a function of $x$, as in Figure 10,
with the fastest-growing modes shown as solid lines and the
corresponding zero-crossing modes shown as dashed lines. The
differences between the zero-crossing solutions and the $x(\rthin)=x_o$
modes is even smaller than in Figure 10; the peak in $\beta(x)$ is
displaced slightly outward and $\beta$ declines smoothly to zero with
increasing $x$. For the $\delta >1$ modes, which peak well away from
the origin, the width of the peak in $\beta(x)$ at half-maximum is
negligibly different in the two cases. For $\delta\lessapprox 0.75$, it is
evident that the difference between the zero-crossing modes and the
optically-thin-boundary modes is negligible for all growth rates.

\subsection{X-Ray Binaries}
In X-ray binary systems the situation is rather different. It is
likely that the accretion disk remains optically thick to the physical
outer boundary of the disk. However, the disk will again not obey
equation (10) throughout, because the assumption of a power-law
surface density cannot hold for the entire disk. What we have taken to
be the outer boundary of the disk in the zero-crossing solutions
corresponds to the circularization radius, \rcirc, in a real X-ray binary
system. The disk will actually extend to the tidal truncation radius \rout,
which is larger by a factor of $2-3$ for reasonable mass ratios (\eg
\cite{fkr92}). In this outer disk region, $\rcirc < R\le\rout$, the
radial velocity $V_R=0$, since there is no flux of mass through this
region of the disk, only angular momentum. Conservation of angular
momentum then requires that
\be
\nu_1\Sigma=\left(\nu_1\Sigma\right)_\rcirc \left({\rcirc\over
R}\right)^{1/2}\;.
\ee
To determine the surface density distribution in the outer disk thus
requires an assumption about the viscosity. We assume that $\nu_1=\nuout$ is
constant in the outer disk; in fact, since \rout\ is only $\sim
2-3\rcirc$, the results are entirely insensitive to this assumption
unless $\nu$ is an extremely strong function of radius in the outer
disk. The term in brackets on the LHS of equation (1) is then
\be
\left[V_R-\nu_1{\Omega'\over\Omega}-{1\over 2}\nu_2
{\left(\Sigma R^3 \Omega\right)^\prime\over \Sigma R^3 \Omega}\right]
={1\over 2}{\nuout\over R}\left(3-\eta\right)\;.
\ee
Fourier transforming with respect to time and linearizing as before,
the twist equation becomes
\be
\eta W''+\left[{\eta-3\over R}-{2i\Gamma_{\rm out}\over\nuout}\right]W' -
{2i\sigma W\over\nuout}=0
\ee
where $\Gamma_{\rm out}$, defined as in equation (5), is constant
since $\Sigma R^2\Omega$ is also constant under the assumption that
$\nu$ is independent of radius in the outer disk. This does not
include the torque due to the companion star, which can be of vital
importance for the disk modes in X-ray binary systems: as shown in
Maloney \& Begelman (1997b), the companion torque allows retrograde
modes as well as prograde modes to exist. Since the orbital period
is short compared to the viscous timescale in the disk, we average
over azimuth and keep only the leading (quadrupole) term in the
companion torque. This contributes a term $-i\omega_o (R/R_o)^{3/2} W$
to the lefthand side of equation (21), where $\omega_o$ is the
quadrupole precession frequency at the fiducial radius $R_o$ (Maloney
\& Begelman 1997b). 

By continuity, the viscosity \nuout\ and the radiation torque
parameter $\Gamma_{\rm out}$ in the outer disk are given by
\be
\nuout\equiv\nu_1(\rcirc)=\nu_o\left(\rcirc/R_o\right)^\delta
\ee
\be
\Gamma_{\rm out}\equiv\Gamma_o R_{\rm circ}^{\delta-1/2}\;.
\ee
Transforming to radius variable $x$ and nondimensionalizing both
$\sigma$ and $\omega_o$ by $2\eta^3R_s^2/\nu_o\epsilon^4$ (\cf
equation [9]) the linearized twist equation for the outer disk becomes
\be
xW''+\left[{2(\eta-3)\over\eta}-1\right]W'-{ix^2\over\xcirc}W'
={ix^3\over x_{\rm circ}^{2\delta}}\left(\tsigma +\tomegao x^3\right)W\;.
\ee
For typical neutron star X-ray binary parameters, $\xcirc\approx 30$
(assuming $\epsilon\approx 0.1$), while $x_{\rm out}\simeq \sqrt 3 \xcirc$.

The value of $W$ at the outer boundary is arbitrary, but we require
that $W'=0$ at \rout; this implies that, in the absence of
tidally-induced precession, there is no torque acting on the outer
disk boundary. As in the case of the AGN (optically thin) boundary
conditions discussed above, the key question is how the outer disk
solutions couple to the inner disk, where equation (10) is valid. In
Appendix E, we show that at \rcirc, the disk must satisfy a jump
condition\footnote[1]{This jump condition is derived by assuming that
mass and angular momentum are added via the accretion stream at a
fixed radius. It assumes {\it no} viscosity discontinuities due to the
stream, and hence is itself an idealization. On the other hand, assuming
the warp goes to zero at \rcirc\ is in some sense
the opposite limit, as it implicitly posits that the effect of
viscosity is to pin the disk to the orbital plane at the
circularization radius. That these two extreme views yield comparable
results is an indication of the insensitivity of our results to the
exact choice of boundary conditions}, given by
\be
\rcirc{\Delta W'\over W}={3\over\eta}
\ee
where $\Delta W'=W'_+-W'_{\_}$ is the jump in $W'$ at \rcirc\
(\ie the difference in the values of $W'$ at radii infinitesimally
larger and smaller than \rcirc). This result was first derived by
J.E. Pringle (1997, private communication). Note that equation (25)
indicates that $W'_+$ is larger than $W'_{\_}$; thus if $W'_+ < 0$,
then $W'_{\_}$, the gradient of $W$ just interior to \rcirc, must be
steeper (more negative).

The outer disk equation can be solved numerically to find the
appropriate boundary condition at \rcirc, and then determine the
matching inner disk solution, as in \S 4.1. However, we can show more
directly that the solutions will always be close to the zero-crossing
solutions. It is straightforward to find the {\it gradient} of the
asymptotic solution for the outer disk, in the usual WKB approximation
(\eg \cite{ben78}) that $W=e^S$ (see also Appendix A):
\be
x{W'\over W}={\left(5+i\chi x\right)\over 2}\pm {i\over 2}\left[\chi^2
x^2-25-ix^2\left\{{10\over\xcirc}+4\chi^2 x_{\rm
circ}^{2-2\delta}\left(\tsigma+\tomegao x^3\right)\right\}\right]^{1/2}
\ee
where $\chi\equiv x/\xcirc \ge 1$. Comparison with numerical solutions
of the outer disk equation shows that equation (26) (with choice of
minus sign) is generally accurate to 10\%, with the worst error being
$15-20\%$ (the latter being the case for some of the prograde disk
modes, which tend to have smaller values of \xcirc\ than the
retrograde modes); equation (26) always underestimates the magnitude
of the gradient. (Note that equation [26] does not predict
$W'\rightarrow 0$ as $x\rightarrow x_{\rm out}$, since the assumption
that $S'' \ll (S')^2$ breaks down as $x$ approaches $x_{\rm out}$.)
Since we are chiefly interested in equation (26) for deriving the jump
condition at the circularization radius, we take the limit
$x\rightarrow\xcirc$, so that $\chi\rightarrow 1$:
\be
x{W'\over W}={\left(5+i \xcirc\right)\over 2}-{i\over 2}\left[x^2_{\rm circ}
-25-i\left\{10\xcirc +4 x_{\rm circ}^{4-2\delta}\left(\tsigma+
\tomegao x^3_{\rm circ}\right)\right\}\right]^{1/2}\;.
\ee
In Figure 13 we plot the value of the gradient in the disk tilt,
$xW'/W$, just outside the circularization radius, for all of the X-ray
binary warped disk models presented by Maloney \& Begelman (1997). The
solid portions of the curves indicate retrograde modes. In all cases
the gradient is negative and generally large, especially for the
retrograde modes; to satisfy the jump condition, the gradient just
inside \rcirc\ must be even steeper (note that $x\Delta W'/W=6/\eta$). 
Because the gradient is always negative and steep where the inner disk
solution patches on to the outer disk solution, the former must have a
shape close to that of the zero-crossing mode with a zero at
\rcirc. Moreover, the tilt of the outer disk solution always goes
rapidly to zero for $R > \rcirc$. For example, Figure 14 plots the
tilt of the outer disk solutions as a function of radius, for the most
rapidly precessing, steady-state ($\tsigma_i=0$) solutions, for
$\delta=0.75$ and $1.25$. Even for the most slowly-declining solution,
the amplitude of the tilt at \xout\ is less than $1\%$ of the value at
\xcirc.

Changing the outer boundary condition does not allow a range of
solutions to exist, unlike the optically-thin-boundary condition
discussed above. The relevant physical question then becomes, what is
the nature of the unstable mode for a given value of \xcirc\ and
\tomegao? In Figure 15 we show the precession and growth rates for the
unstable modes as a function of \tomegao, for a fixed value of
$\xcirc=30$. The upper half of the figure shows the precession rate
and the lower half shows the growth rate. As expected, at small values
of the quadrupole torque parameter the unstable mode is prograde. With
increasing \tomegao\ the unstable mode switches to retrograde
precession, and the growth rate starts to decline, eventually going to
zero. This is easy to understand physically: for a fixed value of
\xcirc, the effect of increasing \tomegao\ is to make the external
torque more important at \xcirc, until eventually this dominates over
the radiation torque in controlling the precession, and forces the
warping mode to become retrograde. At very large values of the
external torque, the viscous torques produced by the driven
differential precession become too strong compared to the radiation
pressure torques, and the instability is switched off; thus the growth
rate goes to zero as \tomegao\ approaches this critical value.

There is actually a second branch of retrograde solutions displayed in
Figure 15 (the dashed curve); these have comparable growth rates to
the modes just discussed but precession rates about an order of
magnitude larger. This second branch consists of modes for which the
gradient in the tilt is {\it positive} (and generally large) at the
circularization radius, unlike the solutions displayed in Figures 13
and 14. For these solutions the warp actually peaks in the outer disk
at $R> \rcirc$ and then declines rapidly towards zero. It is not clear
that these solutions will ever occur in real astrophysical disks, as
our treatment of the outer disk ($\rcirc \le R \le \rout$) is much
more of an idealization than that of the inner disk. 

\section{Discussion}
Earlier work on the radiation-driven warping instability discovered by
Pringle (P96; MBP) considered only the isothermal, $\delta=3/2$
case. In this paper we have considered more general power-law disk
density distributions, from the isothermal disk to $\delta=-3/2$,
corresponding to a radiation-pressure supported disk; this spans the
range that is likely to be relevant to astrophysical disks. Although
the shapes of the eigenfunctions do change with decreasing $\delta$,
the most important features of the instability are generic. Most
importantly, the instability exists over the entire range of surface
density index that we have considered, and the critical radius above
which disks are unstable to radiation-driven warping changes only by a
factor of $\simeq 6$ from $\delta=3/2$ to $\delta=-3/2$. Similarly,
the growth and precession rates (in dimensional units) do not depend
strongly on $\delta$ (see 
the discussion after equation [18] and below). Evaluating
equation (15) for the critical radius,
\bea
R_{\rm cr}&=&\left(5.9\times10^8\;\,\quad -\quad 3.5\times
10^9\right)\; \left({\eta\over  
\epsilon_{0.1}} \right)^2\left({M\over\msol}\right)\;{\rm cm} ,
\nonumber \\
&=& \left(5.9\times10^{16}\quad - \quad 3.5\times 10^{17}\right)\,
\left({\eta\over  
\epsilon_{0.1}} \right)^2\left({M\over 10^8 \msol}\right)\;{\rm cm} , 
\eea
where the range in numerical values is for $\delta=3/2$ to
$\delta=-3/2$ and $\epsilon=0.1\epsilon_{0.1}$. The only warping modes
with zeros at $R < R_{\rm cr}$ are damped, so that disks that are
smaller than $R_{\rm cr}$ are stable against warping. In consequence
of the $\epsilon^{-2}$ scaling of $R_{\rm cr}$, accretion disks in
systems with very low radiative efficiency will not be unstable to
radiation-driven warping unless they are implausibly large. For this
reason, this mechanism cannot provide an explanation for the warp in
the thin maser disk of NGC 4258 (\eg \cite{miy95}, \cite{her97}) if
the inner disk is advection-dominated with $\epsilon\sim 10^{-3}$
(\cite{las96}; see the discussion in MBP), since the maser disk would
be far too small for instability in this case. This also indicates
that radiation-driven warping generally will not be important in
cataclysmic variables or protostellar disks dominated by
accretion-powered luminosity, since the radiative efficiency is
limited to small values as the stellar surfaces are at $R_*\gg R_s$
(but see \cite{arm97} for a discussion of the possible action of the
instability in the protostellar case).

To evaluate the typical precession timescales, we need to evaluate the
viscous inflow timescale at $R_{\rm cr}$. Letting $\nu_1=\alpha c_s
H$, where $c_s$ is the isothermal sound speed and $H$ is the scale
height, we can write the viscous timescale as \be t_{\rm visc}\sim
{2\over 3} {R\over V_\phi}\alpha^{-1} (H/R)^{-2} \ee where $V_\phi$ is
the rotational velocity (assumed to be Keplerian) and $H/R$ is evaluated at
the radius in question. Since $t_{\rm visc}\propto R^{3/2}$, and
$R_{\rm cr}/R_o=x_{\rm cr}^2$, 
\be t_{\rm visc}(R_{\rm cr})\sim {1\over
3}\left({\eta\over\epsilon} \right)^3 {R_s\over\alpha c} x_{\rm
cr}^{3/2}\left(H/R\right)^{-2} 
\ee 
where $H/R$ is now evaluated at $R_{\rm
cr}$. Taking the precession timescale $t_{\rm prec}=2\pi/\sigma_r$,
where $\sigma_r$ is given by equation (18), and evaluating the
constants, we find \be t_{\rm prec}\sim 12\,
{\eta^2\over\epsilon_{0.1}^3} {M/\msol\over
\alpha_{0.1}}\left({H/R\over 0.01}\right)^{-2}_{R_{\rm cr}}\;{\rm
days} \ee with only weak dependence on $\delta$: the numerical
coefficient only varies by a factor of two over the whole range of
$\delta$. Thus the precession timescales for X-ray binary systems (the
only systems in which precession can actually be observed) are
expected to be of the order of weeks to months. 

This is of course the precession timescale for the steady-state modes
from linear theory. As discussed in \S 3, real disks will ordinarily
be unwarped beyond some maximum radius, either the physical edge of
the disk or where the disk becomes optically thin. This outer boundary,
which will not in general correspond to the critical radius, will
determine the warp growth rate. We expect that the warp will 
eventually saturate at some amplitude (but see Pringle 1997). Assuming
that the disk does reach a steady state, what will the precession rate
be? There is reason to suspect it may not be very different from the
linear theory result. Figure 6 shows that, except for growth rates
very close to the maximum, the real part of the eigenvalue
$\tilde\sigma_r$, \ie the precession rate, is nearly independent of
the growth rate. In the isothermal case, in fact, $\tilde\sigma_r$ is
independent of $\tilde\sigma_i$. This suggests that, however different
modes may couple in reaching the final state, the precession rate will
be similar to the linear steady-state result.

Implicit throughout this paper has been the assumption that the disks
are optically thick to both absorption and re-emission, so that they
are subject to the radiation-driven warping instability. This
requirement imposes a minimum mass accretion rate that must be
exceeded for the disk to be optically thick. In Appendix D, we derive
this critical mass accretion rate for three different possible sources
of opacity in astrophysical disks (electron scattering, dust
absorption, and Kramer's opacity) and show that it does not in general
place any significant limitations on occurrence of the instability.

As discussed in \S 3.2, there is one very important systematic change
in the nature of the instability with $\delta$. The difference in the
behavior of the growing modes for $\delta \ge 1$ and $\delta < 1$ is
of fundamental importance for the evolution of disks warped by
radiation pressure. For $\delta \ge 1$, the fast-growing modes all
have their maximum warp (\ie tilt $\beta$) close to the outer edge of
the disk, and the amplitude $\beta_{\rm max}$ is much greater than
$\beta_o$, the tilt at the origin. This immediately implies that the
warp must reach the nonlinear regime when the tilt at small radius is
negligible. In this case the evolution of the disk at radii interior
to the warp maximum is almost certainly driven by the nonlinear
evolution of the outer warp (\eg \cite{pri97}), so that nonlinear
effects {\it must} be important if the warp extends to the disk inner
edge.

For $\delta < 1$, the behavior is qualitatively different, as
$\beta_{\rm max}/\beta_o$ is always of order unity. In this regime,
nonlinearity will be important only if the warp has grown out of the
linear regime at the origin. Furthermore, because the shapes of the
growing warps in these two regimes are so dissimilar, the effects of
shadowing of the central source by the warping of the disk will be
very different. These distinctions are liable to be crucial for X-ray
binaries such as SS 433 and Her X-1, which show evidence for a {\it
global} precessing warp.

One final point regarding X-ray binary systems must be mentioned. In
one of the best-studied systems, Her X-1, the direction of precession
of the warp is inferred to be retrograde with respect to the direction
of rotation (\eg \cite{ger76}) and this has also been suggested for SS
433 (\cite{leib84}; \cite{bri89}). As shown in Appendix B, in the
absence of external torques the direction of precession of the warp
must be prograde. However, the qualification on this statement is
extremely important: as pointed out in \S 4.2, and discussed in detail
by Maloney \& Begelman (1997b), including the quadrupole torque from a
companion star allows retrograde as well as prograde solutions to
exist.

The zero-crossing outer boundary condition that we have imposed will
not be strictly correct in real astrophysical disks. However, as
discussed in \S 4, the solutions that obey the likely realistic outer
boundary conditions -- the optically thin outer boundary for accretion
disks in active galactic nuclei, and a flat outer boundary for disks
in X-ray binaries -- are in all important respects similar to the
zero-crossing solutions.

Radiation-driven warping and precession offers a robust mechanism for
producing tilted, precessing accretion disks, in accreting binary
systems such as Her X-1 and SS 433, and in active galactic
nuclei. Because radiation-driven warping is an inherently global
mechanism, it avoids the difficulties inherent in other proposed
mechanisms for producing warping and precession, \eg communicating a
single precession frequency through a fluid, differentially-rotating
disk. This mechanism can thus explain the simultaneous precession of
inner disks (as evidenced by the jets of SS 433 and the pulse profile
variations of Her X-1) and outer disks (as required to match the
periodicities in X-ray flux and disk emission in these same objects).

A full understanding of the nature of the radiation-driven warping
instability will require nonlinear simulations of the type
presented in Pringle (1997), which will not only allow for inclusion
of the nonlinear terms but also inherently nonlinear effects such as
shadowing. This will be the subject of future work.

\acknowledgements We have greatly benefited from discussions with and
comments from Jim Pringle and Phil Armitage, who also kindly provided
results in advance of publication. We are especially grateful to Jim
Pringle for his insightful comments regarding the choice of outer
boundary conditions. We would also like to thank the referee for
helpful comments on the paper. PRM was supported by the NASA Long
Term Space Astrophysics Program under grant NAGW-4454. MCB acknowledges
support from NSF grant AST-9529175. The research of PRM and MCB was
supported by NASA grant NAG5-4061 under the Astrophysical Theory
Program. MAN was supported by the NASA LTSA Program under grant
NAG5-3225.

\appendix

\section{Asymptotic Solutions to the Twist Equation}
In this appendix we examine the behavior of the twist equation at
large $x$, and show that, unlike the $\delta=3/2$ case studied by MBP,
the solutions for $W$ always diverge as $x\rightarrow\infty$. We also
show from the asymptotic solutions that the eigenfunctions for the
case $\delta=1$ always diverge as the growth rate
$-\tsigma_i\rightarrow 0.25$, independent of the real eigenvalue
$\tsigma_r$, as is seen in the numerical solutions (\S 3.2).

We write the twist equation (10) as
\be
xW''+(2-ix)W' -i\tsigma x^\mu W=0
\ee
where primes denote derivatives with respect to $x$ and $\mu\equiv
3-2\delta \ge 0$ for $\delta\le 3/2$; recall that $x$ 
is real and $\tsigma$ is complex. For any value of $\mu$, this
equation has an irregular singular point at infinity. Assuming that
$W=e^S$ (\eg \cite{ben78}), we obtain
\be
xS'' + x\left(S'\right)^2+(2-ix)S'-i\tsigma
x^\mu=0.
\ee
We further assume that $S''\ll \left(S'\right)^2$. As
$x\rightarrow\infty$, the $2S'$ term can be neglected compared to
$-ixS'$; however, we retain it for reasons that will become obvious
below. Equation (A2) is then
\be
\left (S'\right)^2 +\left({2\over x}-i\right)S' -i\tsigma x^\kappa=0
\ee
where $\kappa\equiv\mu-1\ge -1$. The solution to this equation is
\be
S'\sim {1\over 2}\left[\left(i-{2\over x}\right)\pm\left(-{4i\over
x}-1 +4i\tsigma x^\kappa\right)^{1/2}\right]\;.
\ee
This equation has qualitatively different behavior at large $x$ for
$\kappa < 0$ and $\kappa \ge 0$, corresponding to $\delta > 1$ and
$\delta\le 1$, respectively. (This difference in behavior is of course
also seen in the numerical solutions, as discussed in \S 3.) If $\kappa
< 0$, then all the 
$x-$terms in the square root are small compared to unity, and
expansion of the square root gives the two solutions
\be
S'\sim i-{2\over x}+\tsigma x^\kappa,\qquad -\tsigma x^\kappa\;;\qquad
-1\le\kappa < 0.
\ee
(Note that the validity of the assumption that $S''\ll
\left(S'\right)^2$ requires that $\tsigma x^\kappa \gg \kappa/x$; this
obviously breaks down as $\kappa\rightarrow -1$.)
Integrating and exponentiating then gives the two solutions for $W$
\bea
W_+&\sim& e^{ix} x^{-2}e^{\left(\tsigma/\mu\right)x^\mu},\qquad 
W_-\sim e^{-\left(\tsigma/\mu\right)x^\mu}\;;\qquad 0 < \mu < 1 \\
W_+&\sim& e^{ix} x^{\tsigma-2},\qquad W_-\sim x^{-\tsigma}\;;\qquad
\hskip 1.05truein\mu=0.
\eea
The general solution will be a linear combination of $W_+$ and
$W_-$. Equation (A7) is just the asymptotic behavior of the Kummer
functions $M(a,b,x)$ (\cf MBP; \cite{abs64}, eq. [13.5.1]), with
$a=\tsigma$ and 
$b=2$. Equation (A7) shows that for $\mu=0$ we can impose the condition
$W\rightarrow 0$ as $x\rightarrow\infty$ provided that $0 < \tsigma_r
< 2$. However, equation (A6) shows that it is not possible to impose
this condition for $0 < \mu < 1$ if $\sigma_r\neq 0$: one solution for
$W$ always diverges as $x\rightarrow\infty$. Thus, unlike the
degenerate $\delta=3/2$ solutions, $W$ always diverges as
$x\rightarrow\infty$, for any value of $\tsigma_r$.

For $\kappa=0$, corresponding to $\delta=1$, the $\tsigma-$term under
the square root in equation (A4) is independent of $x$. Since
$|4i\tsigma|$ is not necessarily small compared to unity, we expand
the square root as
\be
i\left(1-4i\tsigma+{4i\over x}\right)^{1/2}\simeq
\omega^{1/2}\left(i-{2\over\omega x}\right)
\ee
with $\omega\equiv 1-4i\tsigma$. We thus obtain
\be
S_\pm\sim {ix\over 2}-\ln x \pm \omega^{1/2}\left(ix-{1\over\omega}\ln
x\right).
\ee
The first term inside the parentheses on the righthand side of (A9)
will cause $W$ to diverge as $x\rightarrow\infty$, as $\exp({\rm
Im}(\omega^{1/2})x)$. Since ${\rm Re}(\omega^{1/2})\rightarrow 0$ as
$\tsigma_i\rightarrow -0.25$, the second term will cause $W$ to
diverge as $\tsigma_i\rightarrow -0.25$, independent of $\tsigma_r$
(see also Appendix C). This divergence is reflected in the behavior of
the zero-crossing solutions; see \S 3.2 and Figures 5a and 5b.

Finally, consider the case $\kappa > 0$. In this case, at large $x$
the $x^\kappa$ term under the square root of equation (A4) dominates,
and so 
\be
S_\pm\sim {ix\over 2}-\ln x\pm{\left(i\tsigma\right)^{1/2}
\over\lambda}x^\lambda
\ee
where $\lambda\equiv 1+\kappa/2=(\mu+1)/2$; $\lambda > 1$ for $\delta
< 1$. For any $\tsigma\ne 0$, the third term causes $W$ to diverge
exponentially as $x\rightarrow\infty$, as for the other solutions for
$\delta < 3/2$.

\section{Proof of the Distinctness of the Eigenvalues}
We separate the twist equation into real and imaginary parts:
\bea
x\beta''+x\beta\gamma'\left(1-\gamma'\right)+2\beta'&=&-x^\mu\beta\tsigma_i\\
x\gamma''+{x\over\beta}\beta'\left(2\gamma'-1\right)+2\gamma'&=&x^\mu
\tsigma_r
\eea
where $\mu\equiv 3-2\delta$ and primes denote derivatives with respect
to $x$. Equation (B2) shows that a necessary condition for an
eigenvalue to occur is $2\gamma'-1=0$ (\ie $\gamma'\rightarrow 1/2$ as
$\beta\rightarrow 0$). Multiplying equation (B2) by $2x\beta^2$ and
rearranging, we obtain
\be
\left[x^2\beta^2\left(2\gamma'-1\right)\right]'=2x\beta^2 
\left(x^\mu\tsigma_r-1\right)\;.
\ee
Integration then gives
\be
x^2\beta^2\left(2\gamma'-1\right)=2\int_0^x\beta^2x\left(x^\mu\tsigma_r
-1\right)\;dx\;.
\ee
Consider first the case $\mu > 0$ ($\delta < 3/2$). Changing variables
to $\omega=\tsigma_r x^\mu$, we rewrite equation (B4) as
\be
\left(2\gamma'-1\right)={2\over \mu\omega^{2/\mu}\beta^2}
\int_0^\omega (\omega-1)\omega^{2/\mu-1}\beta^2\;d\omega\;.
\ee
The integral on the righthand side of (B5) is monotonic; hence, there
can be at most one zero for a given $\tsigma_r$. Note also that
$\tsigma_r$ must be greater than zero for there to be an eigenvalue,
\ie the precession of the warp {\it must} be prograde (in the same direction
as the rotation of the disk).

The change of variables to $\omega$ is not valid for $\mu=0$
($\delta=3/2$). In this case equation (B3) is simply
\be
\left[x^2\beta^2\left(2\gamma'-1\right)\right]'=2x\beta^2 
\left(\tsigma_r-1\right)\;.
\ee
If $\tsigma_r=1$, the righthand side is identically zero, and so
\be
\gamma'={1\over 2}+{C\over 2x^2\beta^2}
\ee
where $C$ is a constant. The boundary condition
$\gamma'\rightarrow\tsigma_r/2$ as $x\rightarrow 0$ for $\delta=3/2$
(\cf equation [12] in the text) requires that $C=0$, and thus for
$\tsigma_r=1$, $\gamma'=1/2$ for all $x$, as can be seen from the
Kummer $M$ functions with $a=\tsigma_r=1$ (MBP).

If $\tsigma_r\ne 1$, we integrate equation (B6) to obtain
\be
\left(2\gamma'-1\right)={2\over x^2\beta^2}\left(\tsigma_r-1\right)
\int_o^x\beta^2 x \;dx\;.
\ee
This equation shows that there are no zeros for $\tsigma_r\ne 1$, as
the integral on the righthand side is positive definite. Instead we
simply have $\gamma'> 1/2$, $\tsigma_r > 1$; $\gamma' < 1/2$,
$\tsigma_r < 1$. 

\section{Solution for the case $\delta=1$}
The twist equation (10) can be re-written as
\be
{\partial\over\partial x}\left[x^2 e^{-ix}{\partial W\over \partial
x}\right ] +x^2 e^{-ix}\left [ix^{2-2\delta}\tsigma W\right] =0
\ee
which for $\delta=1$ simplifies to 
\be
{\partial\over\partial x}\left[x^2 e^{-ix}{\partial W\over \partial
,x}\right ] +x^2 e^{-ix}\left [i\tsigma W\right] =0 .
\ee
Define the new function $\phi$ by
\be
W=x^{-1} e^{ix/2}\phi
\ee
so that
\be
{\partial W\over \partial x}=x^{-1}e^{ix/2}{\partial\phi\over\partial
x} + x^{-1}e^{ix/2}\phi\left[{i\over 2}- x^{-1}\right]\;.
\ee
Then the twist equation (C2) can be re-written as 
\be
{\partial^2\phi\over\partial x^2}+\left[\left({1\over
4}-i\tsigma\right) +{i\over x}\right]\phi=0\;.
\ee
If we now define $b^2=-(1-4i\tsigma)^{-1}$ and a new radial variable
$y=x/b$, equation (C5) becomes
\be
{\partial^2\phi\over\partial y^2}+\left[-{1\over4}+{ib\over y} 
\right]\phi=0\;. 
\ee
This is Whittaker's equation (\eg \cite{abs64}, equation [13.1.31]),
\be
{\partial^2\omega\over\partial z^2}+\left[-{1\over4}+{\kappa\over z}+
{\left({1\over 4}-\mu^2\right)\over z^2}\right]\omega = 0 
\ee
where the Whittaker functions $\omega= M_{\kappa,\mu}$ are related to
the Kummer functions $M(a,b,z)$ by
\be
M_{\kappa,\mu}= e^{-{z\over 2}} z^{\mu+{1\over 2}}M\left({1\over
2}+\mu-\kappa,1+2\mu,z\right)
\ee
where in our case $\mu=1/2$, $\kappa=ib$. In terms of $W$ and $x$,
this solution is
\be
W(x)=e^{-{(1-ib)\over 2b}x}M\left(1-ib,2,{x\over b}\right)\;.
\ee

\section{Critical Accretion Rate}
For an accretion-fueled source, neither the growth/precession rates
nor the critical radius for instability explicitly depends on the mass
accretion rate, but only on the radiative efficiency $\epsilon\equiv
L/\dot M c^2$. However, the requirement that the disk be optically
thick to absorption and re-emission of the incident flux does impose a
minimum value for the surface density $\Sigma\propto \dot M/\alpha$ in the
usual $\alpha-$viscosity formalism. 

We define a critical mass accretion rate $\dot M_{\rm cr}$ such that
the disk is optically thick at the critical radius for instability,
\ie the surface density $\Sigma=\Sigma_{\rm cr}$ there. We consider
three possible sources of opacity:
\begin{itemize}
\item[(1)] Dust opacity. This will dominate if the gas is molecular or
atomic with $T\lessapprox 10^4$ K. (The dust will be considerably
cooler than the gas.) In this case the Rosseland mean opacity coefficient
$\kappa_R=0.7$, $2.7$, $5.0$, and $7.5$ cm$^2$ gm$^{-1}$ for dust
temperatures $T_d=50$, 100, 300, and 600 K, respectively
(\cite{pol94}; this assumes Solar neighborhood abundances and
depletion patterns). The corresponding critical surface densities are
\be
\Sigma_{\rm cr}({\rm dust})\simeq 4.3\times
10^{23}\kappa_R^{-1}\psqcm\simeq 6\times 10^{22}-6\times 10^{23}\psqcm
\ee
for $T_d=50-600$ K.
\item[(2)] Electron scattering. Ignoring Klein-Nishina corrections, in
this case $\kappa_R\simeq 0.35$ cm$^2$ gm$^{-1}$, and so
\be
\Sigma_{\rm cr}\simeq 1.2\times 10^{24}\psqcm
\ee
independent of $n$ or $T$.
\item[(3)] Kramer's (bound-free and free-free) opacity for ionized
gas. From \cite{schw58} the Rosseland mean opacity in this case is
approximately $\kappa_R\approx 3\times 10^{23}\rho T^{-7/2}\;{\rm
cm^2\;gm^{-1}}$ 
for electron temperatures $T\gtrapprox 10^6 K$, which gives a critical
density
\be
\Sigma_{\rm cr}\approx 6\times 10^{23} n_H^{-1}T^{7/2}\psqcm .
\ee
The numerical coefficient is for $T\sim 10^6$ K; it slowly decreases
with increasing $T$. 
\end{itemize}

From the definition of $R_o$ (\cf equation [8]) we have
\be
{R_{\rm cr}\over R_o}=x_{\rm cr}^2
\ee
and so the requirement that the disk be optically thick at the
critical radius then becomes
\be
\Sigma_{\rm cr}=\Sigma_o x_{\rm cr}^{-2\delta}
\ee
which can be written as a constraint on $\Sigma_o=\Sigma(R_o)$:
\be
\Sigma_o=\Sigma_{\rm cr}x_{\rm cr}^{2\delta}.
\ee
The surface density $\Sigma_o=\dot M/3\pi\nu_o$, which in the
$\alpha-$viscosity prescription is 
$\dot M/3\pi\alpha(c_s H)_o$, where the isothermal sound speed $c_s$
and the disk scale height $H$ are evaluated at $R_o$. For a thin disk
with negligible self-gravity, $H\simeq R c_s/V_{\rm orb}$. From the
definition of $R_o$, $V_{\rm orb}(R_o)=(\epsilon/\eta) c$. To adequate
accuracy we can take the sound speed to be $c_s\simeq 13 T_4^{1/2}$ km
s$^{-1}$, where the gas temperature $T=10^4 T_4$ K, and so
\be 
\Sigma_o=3.8\times 10^{-3} {\dot M\over \alpha} \left(T_4
R_s\right)^{-1} \left({\epsilon\over\eta}\right)^3\;.
\ee
where $R_s$ is in units of cm. If we write $\dot M$ in terms of the
Eddington accretion rate 
\be
\dot M_E\equiv {L_E\over c^2}=4.7\times 10^{11} R_s\;{\rm gm\;s^{-1}} ,
\ee
then
\be
\Sigma_o=7.7\times 10^{32}{\dot m\over\alpha} T_4^{-1}
\left({\epsilon\over\eta}\right)^3\psqcm
\ee
where $\dot m\equiv \dot M/\dot M_E$. Equating this to $\Sigma_{\rm
cr} x_{\rm cr}^{2\delta}$, we finally obtain an expression for the
critical mass accretion rate
\be
{\dot m\over\alpha}=1.3\times 10^{-10}{\Sigma_{\rm cr}\over
10^{23}\psqcm} T_4 \left(\eta\over\epsilon\right)^3 x_{\rm
cr}^{2\delta}.
\ee
Equation (D10) is simply applicable for the case of dust or electron
scattering opacity. However, for Kramer's opacity, we need to know the
density as well as the column density. Solving for the density at
$R_o$ and proceeding as before, we obtain a slightly different
equation for the critical accretion rate:
\be
{\dot m\over\alpha}=7.6\times 10^{-11} T_6^3 R_s^{1/2} 
\left(\eta\over\epsilon\right)^{9/2} x_{\rm cr}^\delta
\ee
where $T_6=T/10^6$ K, a more realistic value for the Kramer's opacity
case.

Consider first equation (D10), for the case of dust and Thompson
opacity:
\begin{itemize}
\item[(1)] The dust temperature is likely to be $T_d\sim\;$a few hundred
K when this is the dominant source of opacity, so that $\Sigma_{\rm
cr}\approx 10^{23}\psqcm$, while $T_4\sim 0.1-1$. For $\delta=3/2$
this is 
\be
{\dot m\over\alpha}=3.2\times 10^{-5} T_4 {\eta^3\over
\epsilon_{0.1}^3}
\ee
where $x_{\rm cr}=2\pi$ and we have normalized
$\epsilon=0.1\epsilon_{0.1}$. For $\delta=1$, the numerical
coefficient is reduced by a factor of 5. This does not impose a
serious constraint on disk {\it stability}; of course, $\dot
m_{cr}\propto x^{2\delta}$, so for $\delta > 0$, opacity effects
become more stringent if the disk is to be optically thick at $R\gg
R_{\rm cr}$ ($x\gg x_{\rm cr}$).
\item[(2)] For electron scattering $\Sigma_{\rm cr}\simeq 1.2\times
10^{24}\psqcm$, while $T\gtrapprox 10^6$ K, so
\be
{\dot m\over\alpha}=1.6\times 10^{-4} T_6 {\eta^3\over
\epsilon_{0.1}^3} x_{\rm cr}^{2\delta}.
\ee
This requirement is most stringent for $\delta > 0$, as $x_{\rm
cr}^{2\delta}$ ranges from one to $(2\pi)^3\simeq 248$ for
$0\le\delta\le 3/2$, although an isothermal disk is unlikely to be a
good assumption in the electron-scattering regime.
\end{itemize}

Now consider equation (D11), for Kramer's opacity.
\begin{itemize}
\item[(3)] In this case $T_6 \gtrapprox 1$, and substituting for $R_s$
in equation (D11)
\be
{\dot m\over\alpha}=1.3\times 10^{-3} T_6^3 \left({M\over
M_\odot}\right)^{1/2} {\eta^{9/2}\over \epsilon_{0.1}^{9/2}}x^\delta
\ee
where $M$ is the mass of the central object. For active galactic
nuclei this can be a significant constraint if $\delta > 0$, since
$M\sim 10^6-10^9\msol$. However, $T_6 \ll 1$ for accretion disks
around AGN at the radii of interest for warping, so this is probably
irrelevant. 
\end{itemize} 

\section{Disk Outer Boundary Jump Condition in X-Ray Binaries}
We start with the basic conservation equations from \cite{pri92} - his
equations (2.1) and (2.2). We assume that matter is injected into the
disk at \rcirc, and that $V_R$ is zero for $\rcirc\le
R\le\rout$. Integrating the mass and angular momentum conservation
equations across \rcirc\ gives the jump conditions
\be
\left.\Sigma V_R\right\vert_{\_}^+={\dot M\over \rcirc}
\ee
\be
\Sigma\left(V_R-\nu_1{\Omega'\over\Omega}\right){\bf l}-{1\over
2}\left.\nu_2 \Sigma {\partial{\bf l}\over\partial
R}\right\vert_{\_}^+={\dot M\over \rcirc} {\bf l_o}
\ee
where {\bf l} is the unit tilt vector giving the direction of angular
momentum at $R$, as before, and ${\bf l_o}$ is the unit vector normal
to the equatorial plane of the binary:
\be
{\bf l_o}=(0,0,1)
\ee
in our usual Cartesian coordinate system. Plus and minus signs denote
quantities evaluated at radii just outside and inside of \rcirc,
respectively. We assume that the disk is continuous, so that ${\bf
l_+}={\bf l_{\_}}={\bf l}$, but that the derivative $\partial {\bf
l}/\partial R$ may be discontinuous across the boundary. Furthermore,
since there may also be a discontinuity in the surface density, we
assume that
\be
\left(\Sigma\nu_1\right)_+=\left(\Sigma\nu_1\right)_{\_}\left({\Sigma_+
\over\Sigma_{\_}}\right)\equiv \left(\Sigma\nu_1\right)_{\_} y\;.
\ee
Substituting from equation (E1) into (E2) and rearranging, we have
\be
\left(y-1\right){\Omega'\over\Omega}{\bf l}+{\eta\over 2}\left[y\left(
{\partial{\bf l}\over\partial R}\right)_+-\left({\partial{\bf
l}\over\partial R}\right)_{\_}\right]={\dot M\over \rcirc\left(\Sigma
\nu_1\right)_{\_}}\left({\bf l}-{\bf l_o}\right)\;.
\ee
Given $\beta'_{\_}$, $\gamma'_{\_}$, and
$\left(\Sigma\nu_1\right)_{\_}$, equation (E5) gives the three
conditions necessary to determine $y$, $\beta'_+$, and
$\gamma'_+$. For example, to find $y$ we take the dot product of
equation (E5) with {\bf l}: since
\be
{\bf l}\cdot\left({\partial{\bf l}\over\partial R}\right)_+=
{\bf l}\cdot\left({\partial{\bf l}\over\partial R}\right)_{\_}=0\;,
\ee
we have
\be
\left(y-1\right){\Omega'\over\Omega}={\dot M\over \rcirc\left(\Sigma
\nu_1\right)_{\_}}\left(1-\cos\beta\right)
\ee
which gives the useful result that in the small angle ($\beta\ll 1$)
limit, $y=1+O\left(\beta^2\right)$, so that the surface density is
continuous in this regime. Similarly, in this limit the
$Z$-component of equation (E5) is also $O\left(\beta^2\right)$, and so
we neglect it. In terms of $W\equiv \beta e^{i\gamma}$, equation (E5)
in the linear (small $\beta$) regime then becomes
\be
{\Delta W'\over W_{\rm circ}}={2\dot M\over \eta
\rcirc\left(\Sigma\nu_1\right)_{\_}}\;.
\ee
Since in the small angle limit we also have
\be
{\dot M\over\rcirc}=-\left(\Sigma V_R\right)_{\_}=-\left(\Sigma
\nu_1\right)_{\_}\left({\Omega'\over\Omega}\right)={3\over 2}{\left(\Sigma 
\nu_1\right)_{\_}\over\rcirc}
\ee
(assuming a Keplerian rotation curve), we finally obtain
\be
\rcirc{\Delta W'\over W_{\rm circ}}={3\over\eta}
\ee
which is the desired jump condition. Equation (E10) was first derived by
J.E. Pringle (1997, private communication).

\clearpage

\figcaption[figure1.eps]{The location of the zero, $x_o$, for the
first ten zero-crossing eigenfunctions, for the steady-state
($\tsigma_i=0$) solutions of the twist equation. The locations of the
zeros are plotted as a function of the surface density power-law index
$\delta$, for $-3/2 \le\delta\le 3/2$. The steps in $\delta$ are
$0.05$. The merging of eigenvalues that occurs for the higher-order
eigenfunctions at $\delta \lessapprox 1$ is marked.}

\figcaption[figure2.eps]{The real eigenvalues for the same
steady-state eigenfunctions as in Figure 1. Plotted are the real
eigenvalues $\tsigma_r$ for the first ten zero-crossing
eigenfunctions. The case $\delta=3/2$ is obviously degenerate
($\tsigma_r=1$ for all order eigenfunctions). The first-order
eigenfunction has the largest eigenvalue for all other values of
$\delta$. The behavior of the higher-order (more distant) zeros
becomes very complex for smaller values of $\delta$, with merging of
the eigenvalues, as seen in Figure 1.}

\figcaption[figure3.eps]{The magnitude of the tilt $\beta$ as a
function of $x$, for the steady-state eigenfunctions. Plotted is
$\beta(x)$ for (left to right) $\delta=1.5$ to $-1.5$, in steps of
$0.25$ in $\delta$. The tilt at the origin has been set to unity.}

\figcaption[figure4.eps]{The location of the zero for the (growing)
time-dependent eigenfunctions. Plotted is $x_o$ for the first-order
growing modes, as a function of the growth rate $-\tsigma_i$. From
left to right, the curves are for $\delta=-3/2$ to $\delta=1.45$, in
steps of $0.05$ in $\delta$. The upper envelope delineates the
maximum growth rates for the first-order modes. For $\delta=1$,
$x_o\rightarrow\infty$ as $-\tsigma_i\rightarrow 0.25$; see Appendices
B and C. The curve for this case has been plotted to
$-\tsigma_i=0.249$.}

\figcaption[figure5.eps]{(a) The location of the zero
as a function of growth rate $-\tsigma_i$, for the first three order
eigenfunctions for $\delta=1$. The zeros merge as
$-\tsigma_i\rightarrow 0.25$ (see Appendices B and C). (b) The real
part of the eigenvalue, $\tsigma_r$, as a function of growth rate
$-\tsigma_i$, for the three $\delta=1$ eigenfunctions of Figure
5a. The real part of the eigenvalue approaches zero as
$-\tsigma_i\rightarrow 0.25$; see Appendices B and C.} 


\setcounter{figure}{5}
\figcaption[figure6.eps]{The real part of the eigenvalue,
$\tsigma_r$, as a function of growth rate $-\tsigma_i$, for the
first-order growing modes shown in Figure 4. For $\delta=1$,
$\tsigma_r\rightarrow 0$ as $-\tsigma_i\rightarrow 0$; for
$\delta=3/2$, $\tsigma_r$ is independent of $\tsigma_i$.}

\figcaption[figure7.eps]{The tilt $\beta$ as a function of radius
variable $x$ for the fastest-growing modes. The maximum value of the
tilt has been normalized to unity. Plotted are (from left to right)
$\beta(x)$ for $\delta =1.45$, $1.35$, $1.25$, $1.15$, and $1.05$. }

\figcaption[figure8.eps]{$\beta(x)$, as in Figure 7, for (from left to
right) $\delta=0.55$, $0.65$, $0.75$, $0.85$, and $0.95$. As in Figure
7, the maximum value of $\beta$ has been normalized to unity.}

\figcaption[figure9.eps]{The real portion of the eigenvalue
$\tsigma_r$ versus the imaginary part of the eigenvalue $\tsigma_i$,
for warping modes that satisfy the optically thin outer boundary
condition discussed in \S 4.1. The optically thin radius has been set
equal to the radius of the first-order zero for the steady-state
solutions for (from left to right) $\delta=1.50$, 1.25, and 0.75. The
dotted portions of the curves have $\tsigma_i > 0$, and are therefore
damped.}

\figcaption[figure10.eps]{The tilt $\beta$ as a function of radius
variable $x$, for the steady-state zero-crossing modes (dotted lines)
and the fastest-growing optically-thin-boundary modes from
Figure 9. The tilt has been normalized to a maximum of unity in all
cases.} 

\figcaption[figure11.eps]{As in Figure 9, but with \rthin\ set equal
to the zero for rapidly-growing zero-crossing modes rather than
steady-state zero-crossing modes.}

\figcaption[figure12.eps]{As in Figure 10, but with \rthin\ set equal
to the zero for rapidly-growing zero-crossing modes rather than
steady-state zero-crossing modes.}

\figcaption[figure13.eps]{The gradient in the disk tilt, $xW'/W$, just
outside the circularization radius, for all of the prograde and
retrograde warp models presented in Maloney \& Begelman (1997b). This
has been derived using the asymptotic solution to the outer disk
equation, given by equation (27). The solid lines indicate the
solution regimes where the precession direction is retrograde; the
prograde solutions are: $\delta=1.25$, $\tsigma_i=0$ (dotted line);
$\delta=0.75$, $\tsigma_i=0$ (short-dashed line); $\delta=0.75$,
maximum $-\tsigma_i$ (long-dashed line); $\delta=1.25$, maximum
$-\tsigma_i$ (dot-dashed line). Note that in all cases the gradient 
is large and negative.}

\figcaption[figure14.eps]{The tilt $\beta$ of the outer disk
solutions, for the most rapidly-precessing, steady-state warp
modes found by Maloney \& Begelman (1997b), for $\delta=0.75$ and
1.25. The retrograde solutions are shown by the solid ($\delta=1.25$)
and dotted ($\delta=0.75$) lines, while the prograde solutions are
shown by the short-dashed ($\delta=1.25$) and long-dashed
($\delta=0.75$) lines.}

\figcaption[figure15.eps]{The steady-state and growing modes for a
fixed value of the circularization radius, $\xcirc=30$. The disk
truncation radius \xout\ is assumed to be $\protect\sqrt3$ times
larger. The precession rate $\tsigma_r$ (top panel) and growth rate
$\tsigma_i$ (bottom panel) are plotted as functions of the companion
torque parameter \tomegao. The dotted curve shows the prograde modes
and the solid curve shows the retrograde modes. The dashed line is a
second branch of retrograde solutions; it is not clear that these will
exist in real accretion disks (see discussion in \S 4.2).}

\clearpage
\plotone{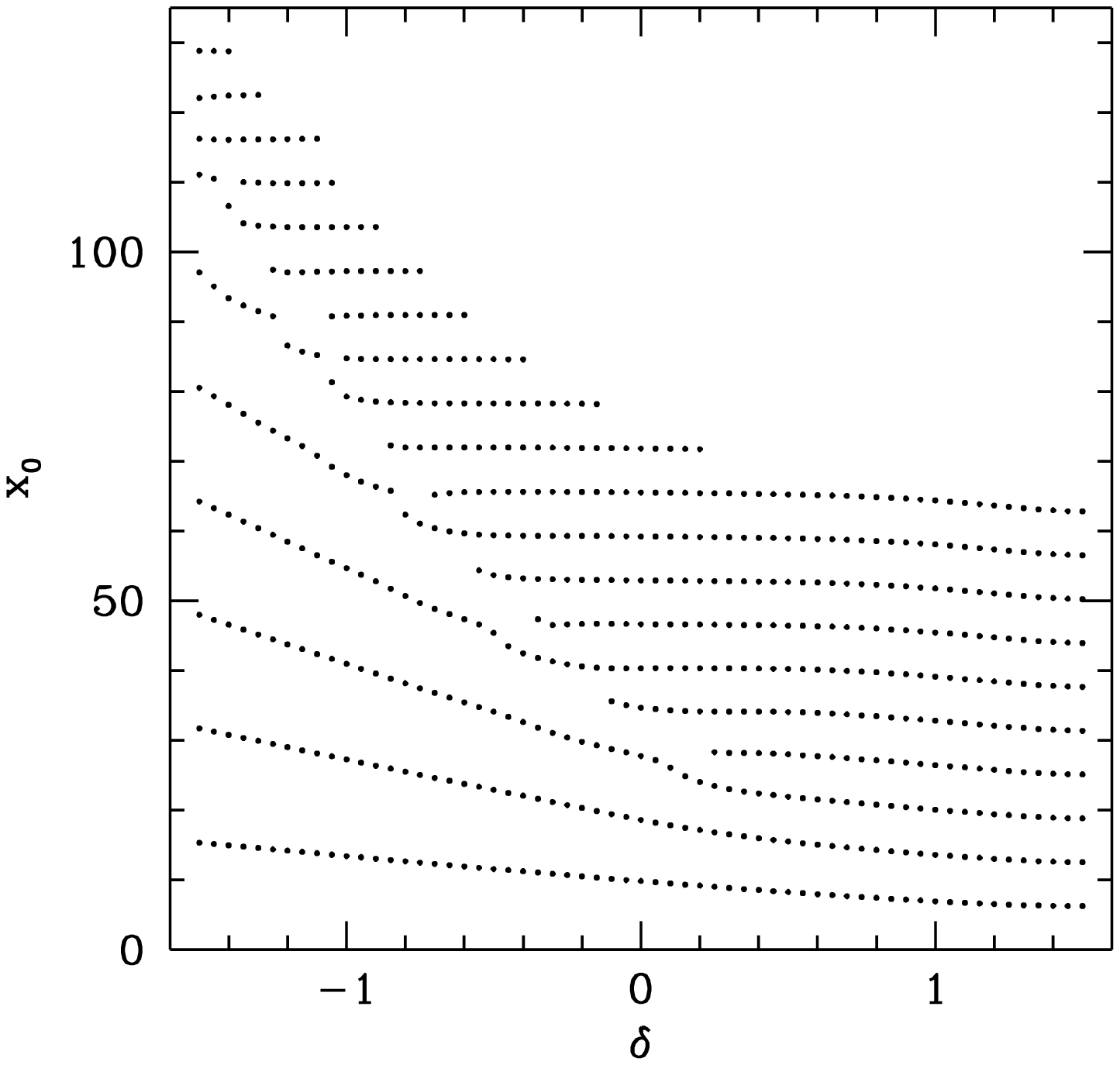}

\clearpage
\plotone{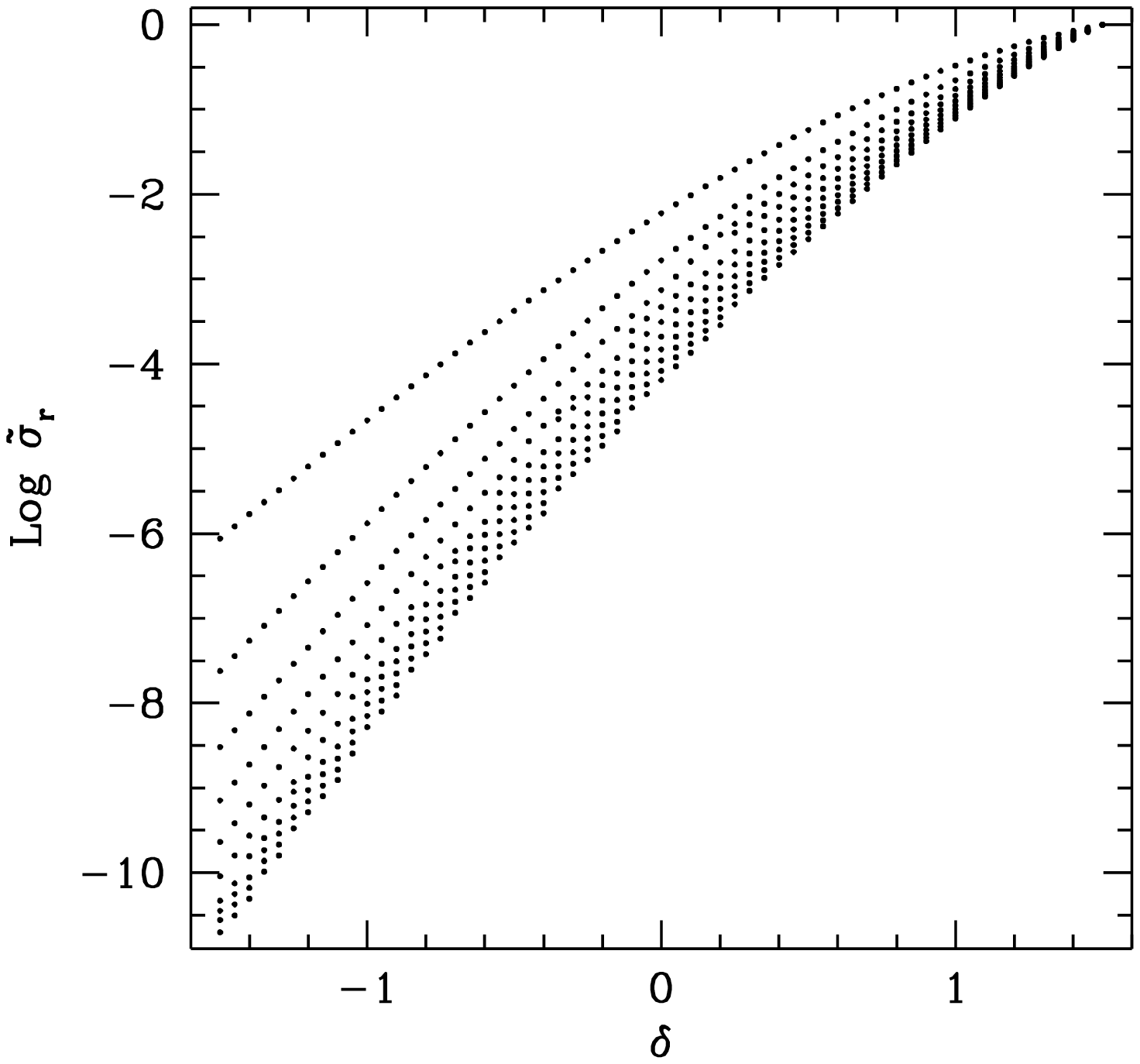}

\clearpage
\plotone{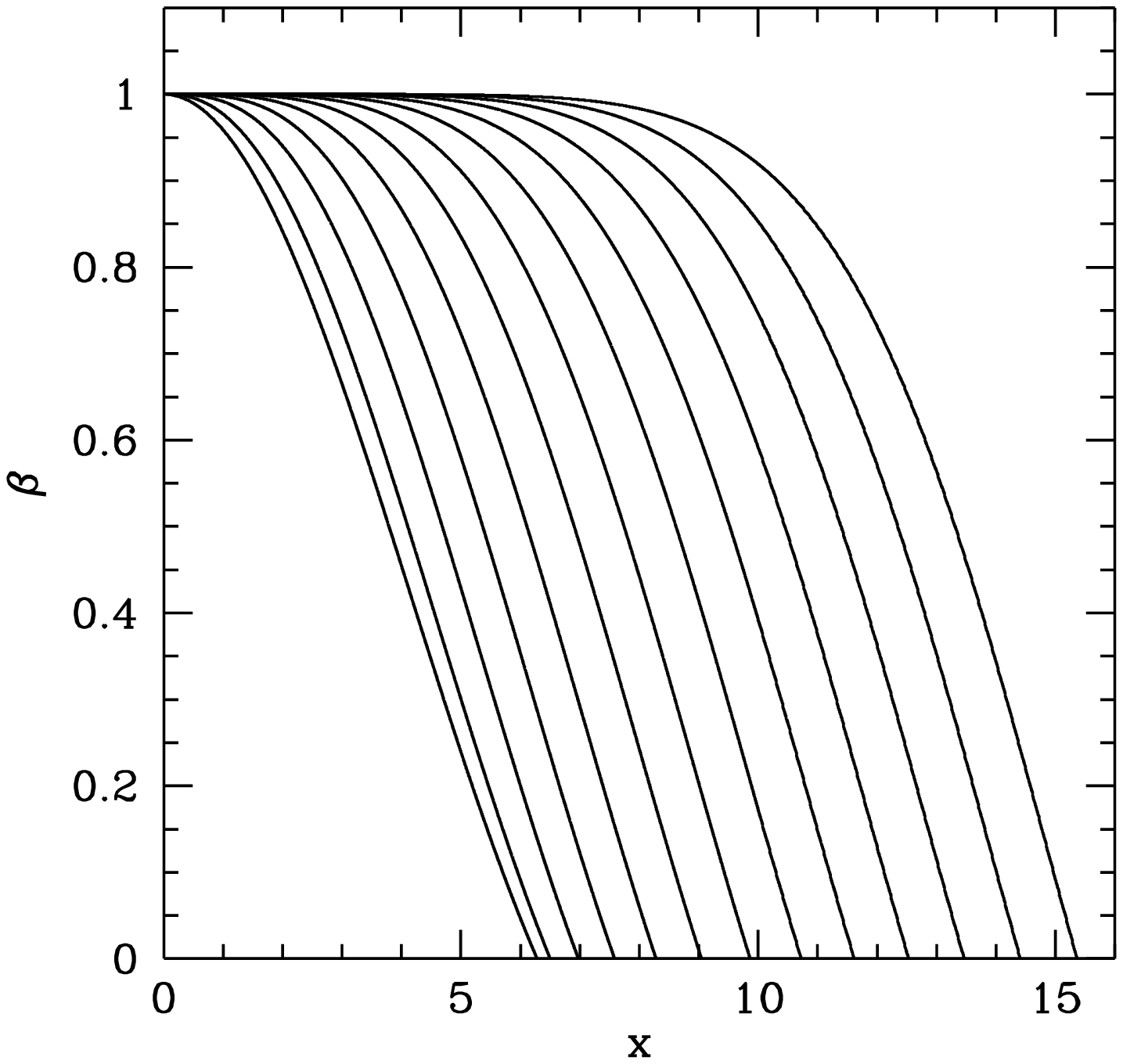}

\clearpage
\plotone{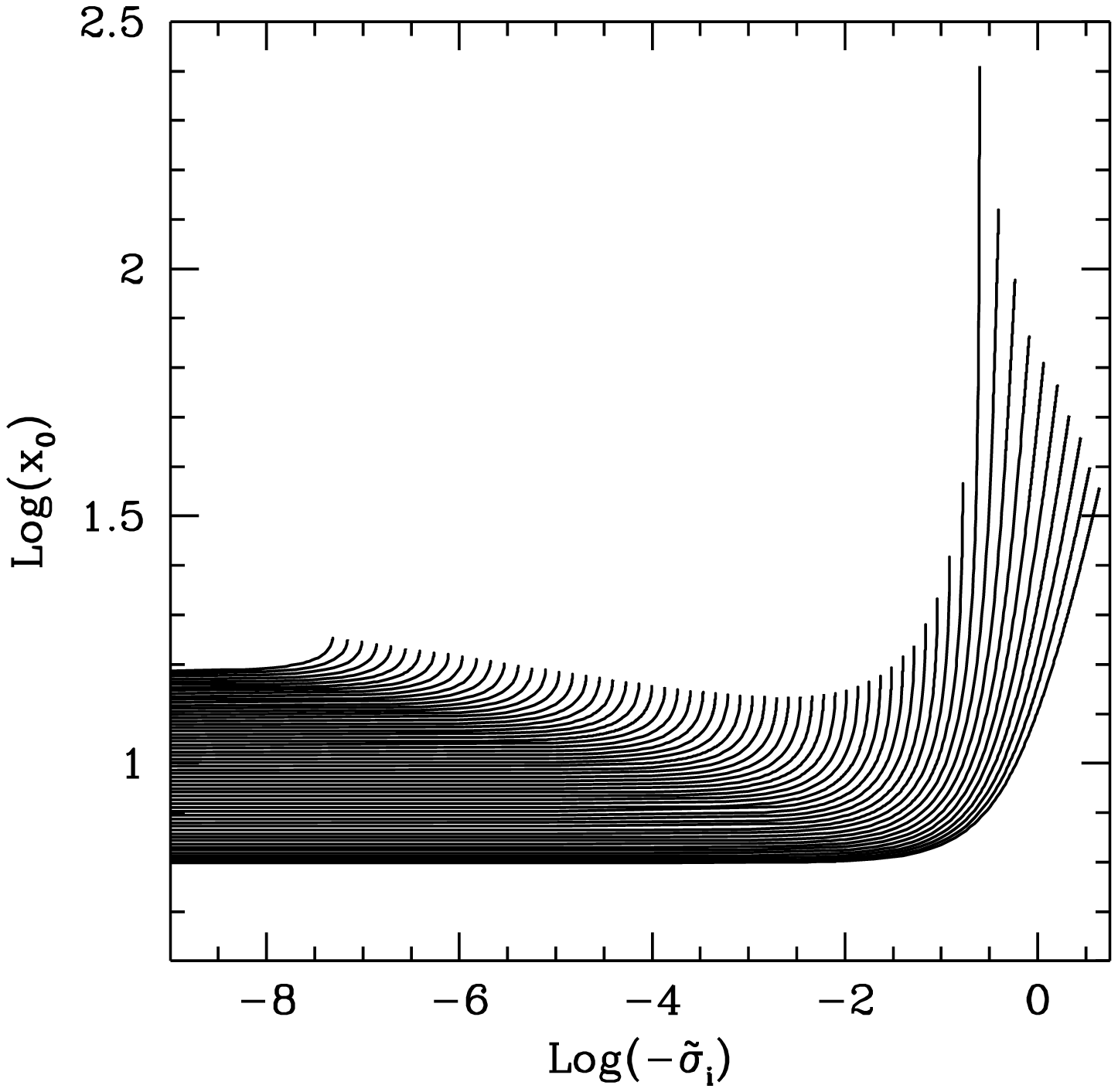}

\clearpage
\plotone{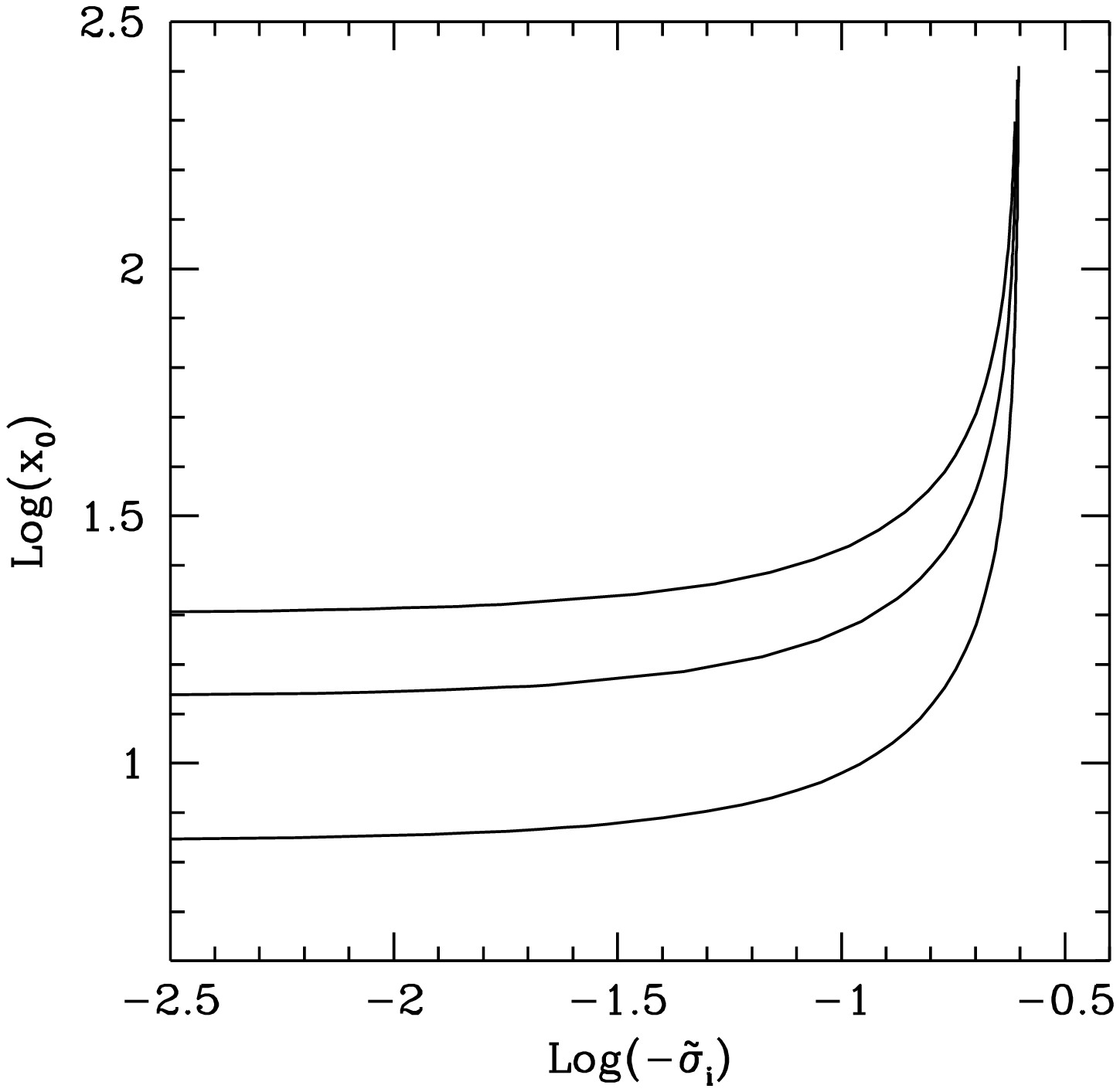}

\clearpage
\plotone{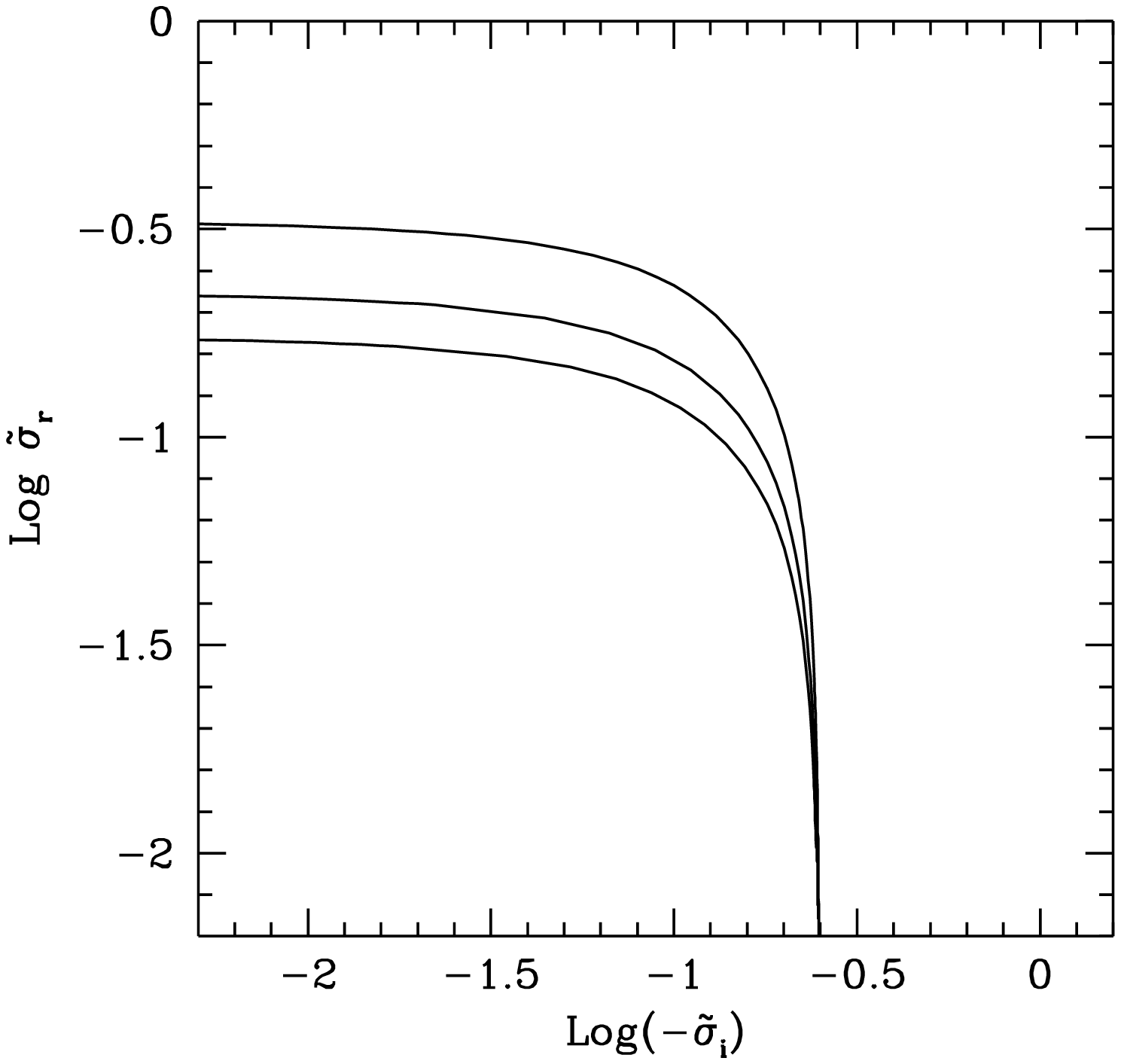}

\clearpage
\plotone{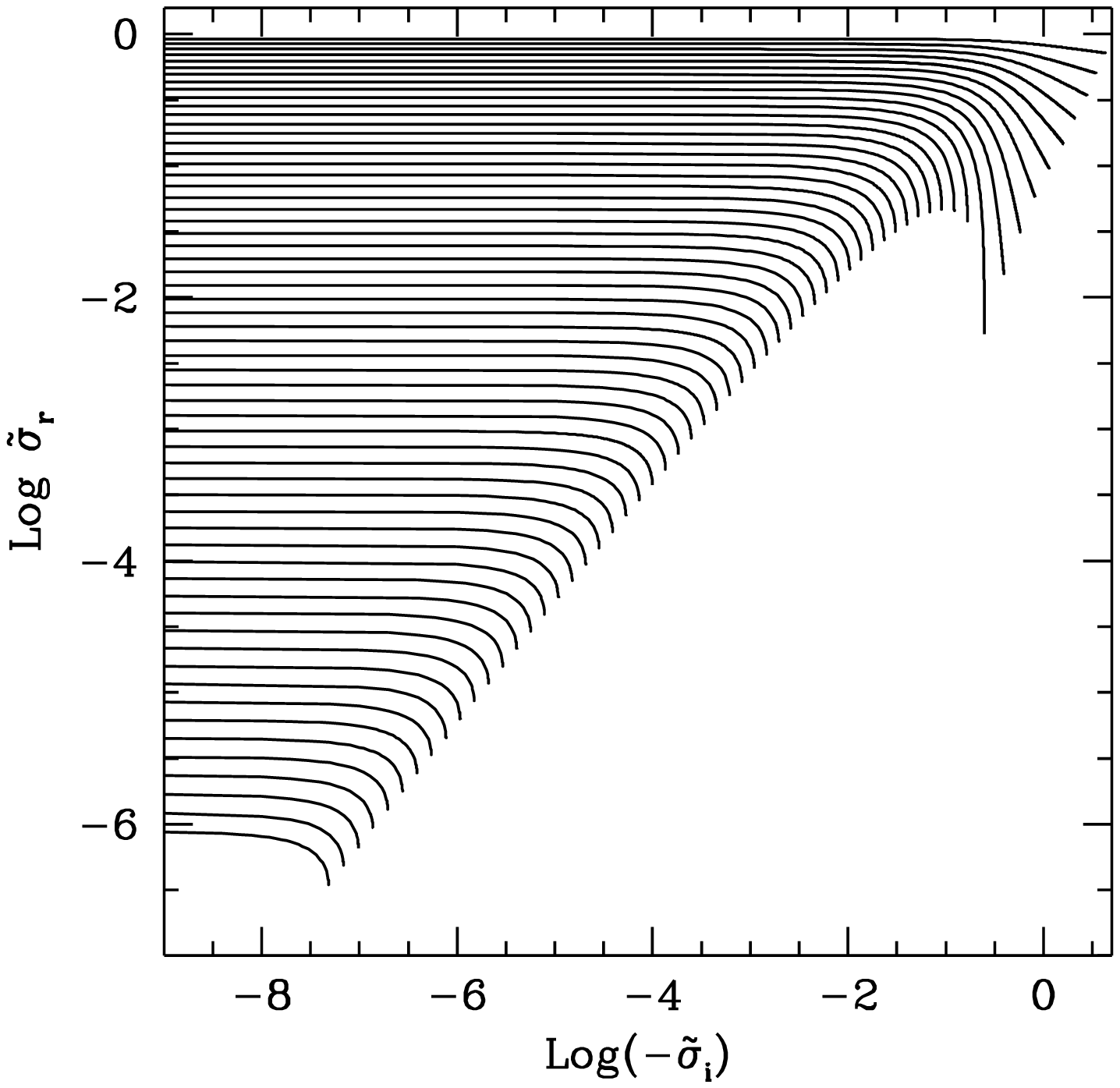}

\clearpage
\plotone{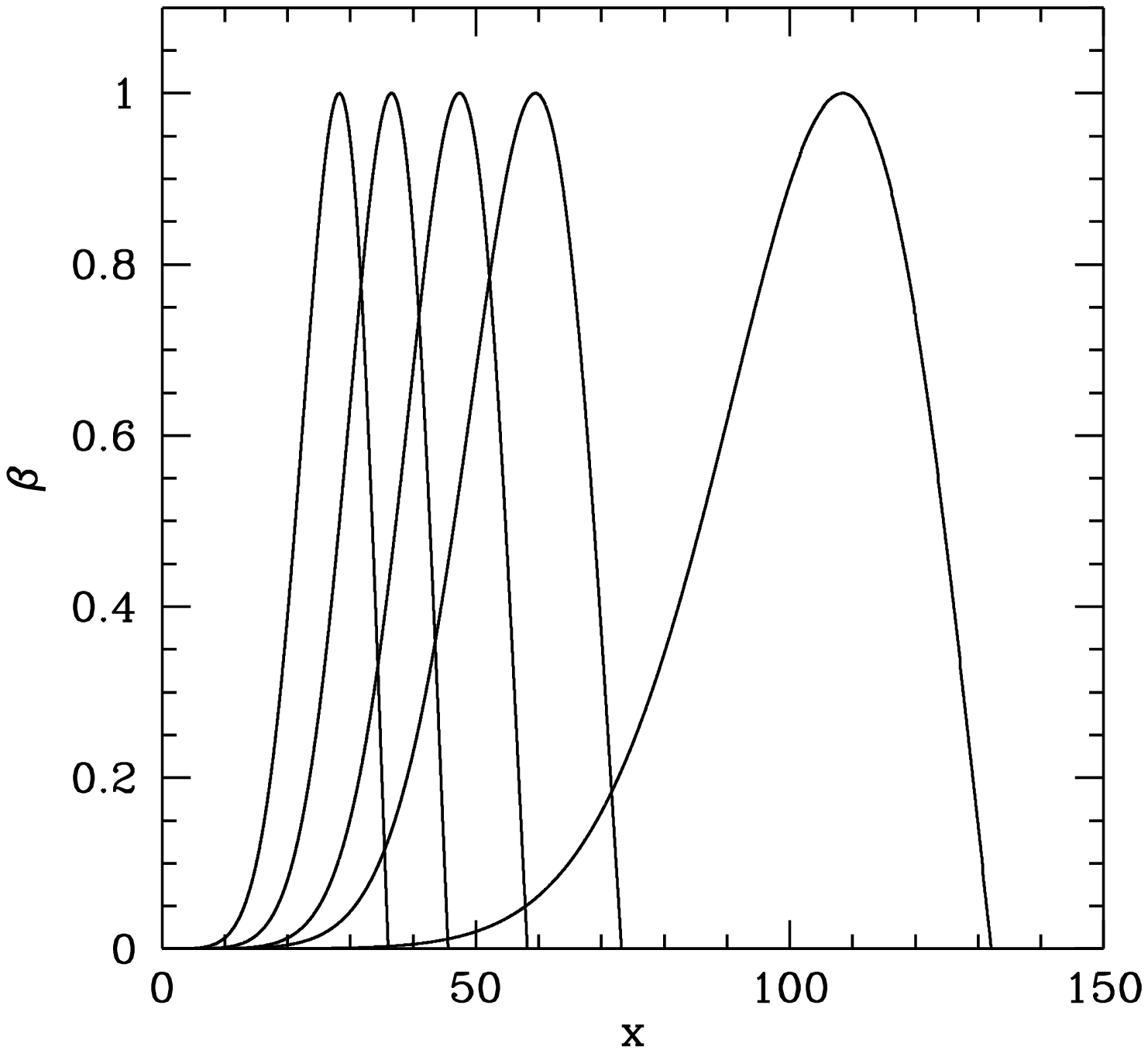}

\clearpage
\plotone{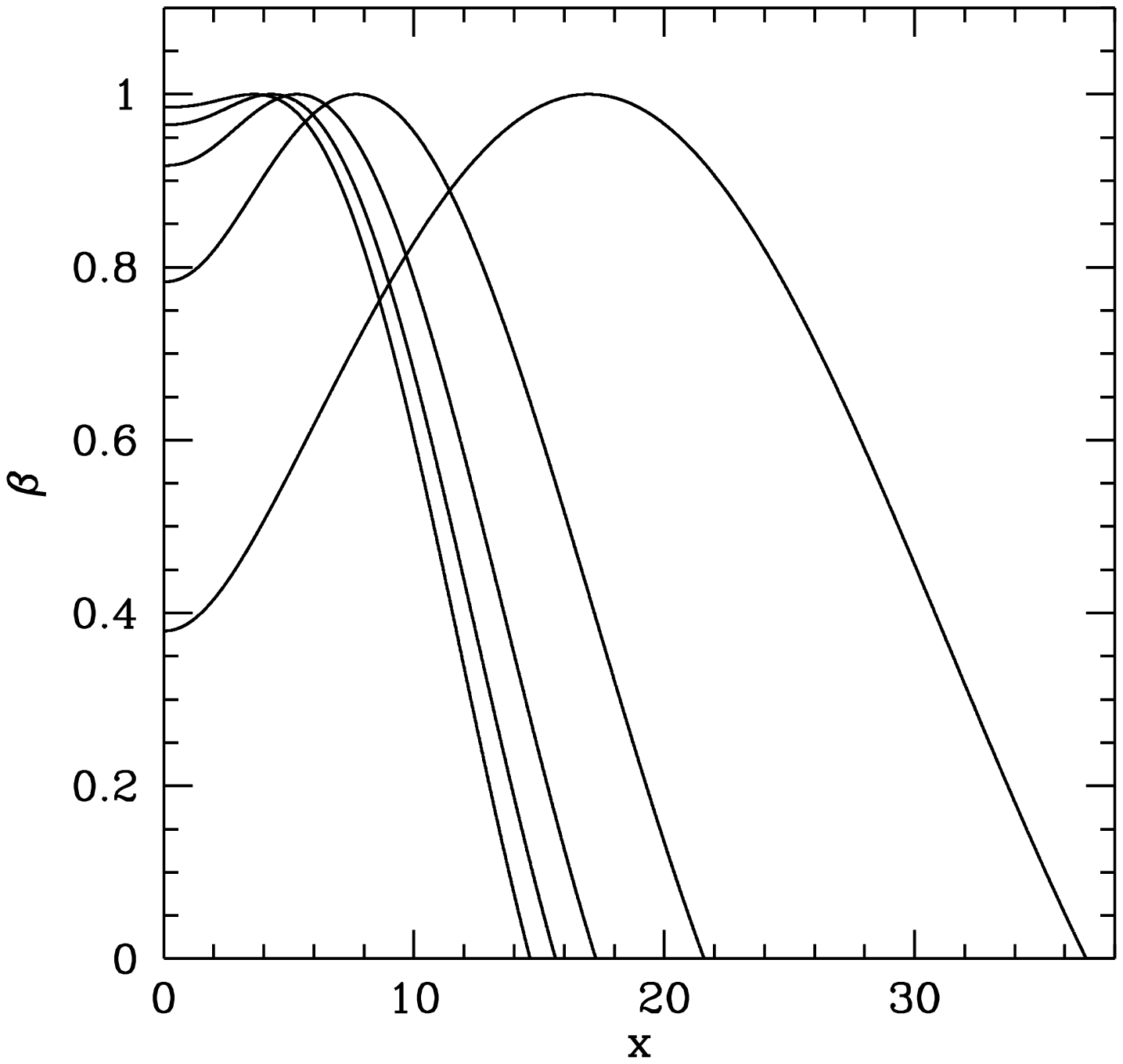}

\clearpage
\plotone{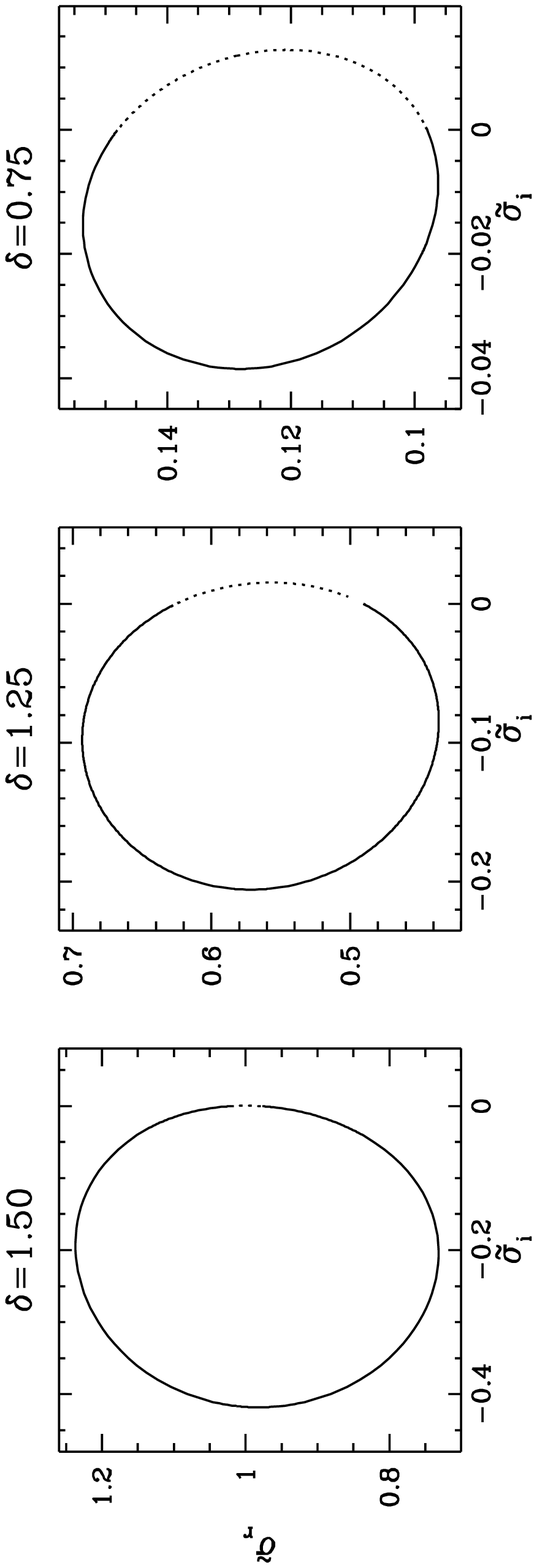}

\clearpage
\plotone{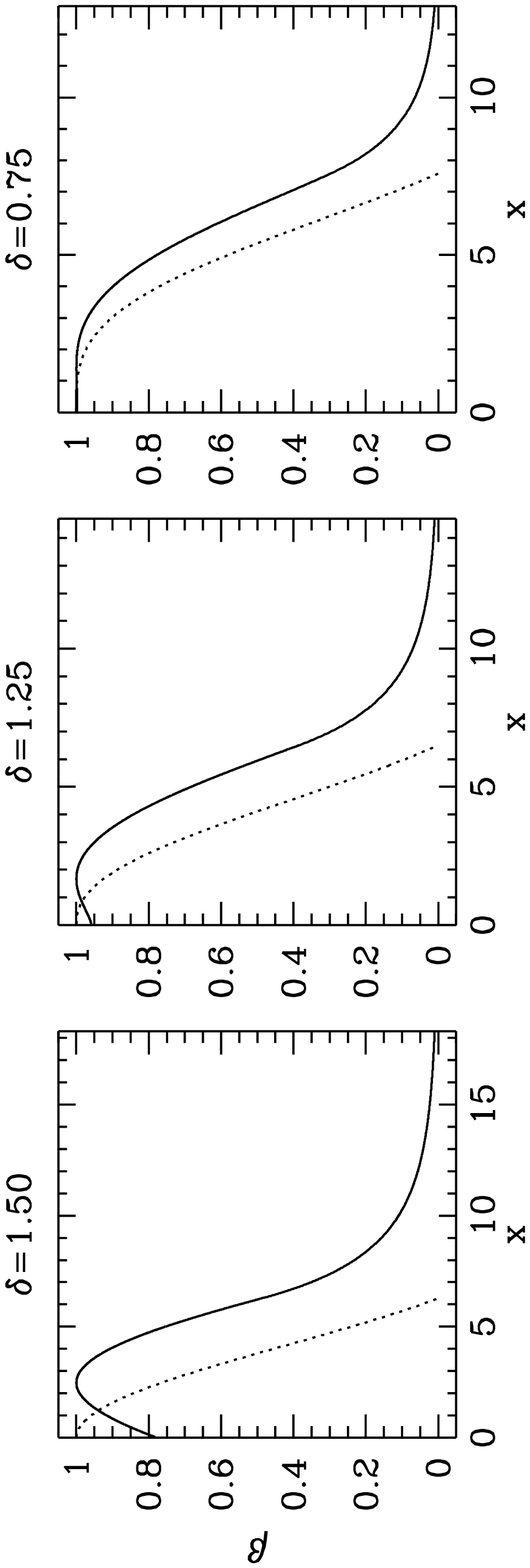}

\clearpage
\plotone{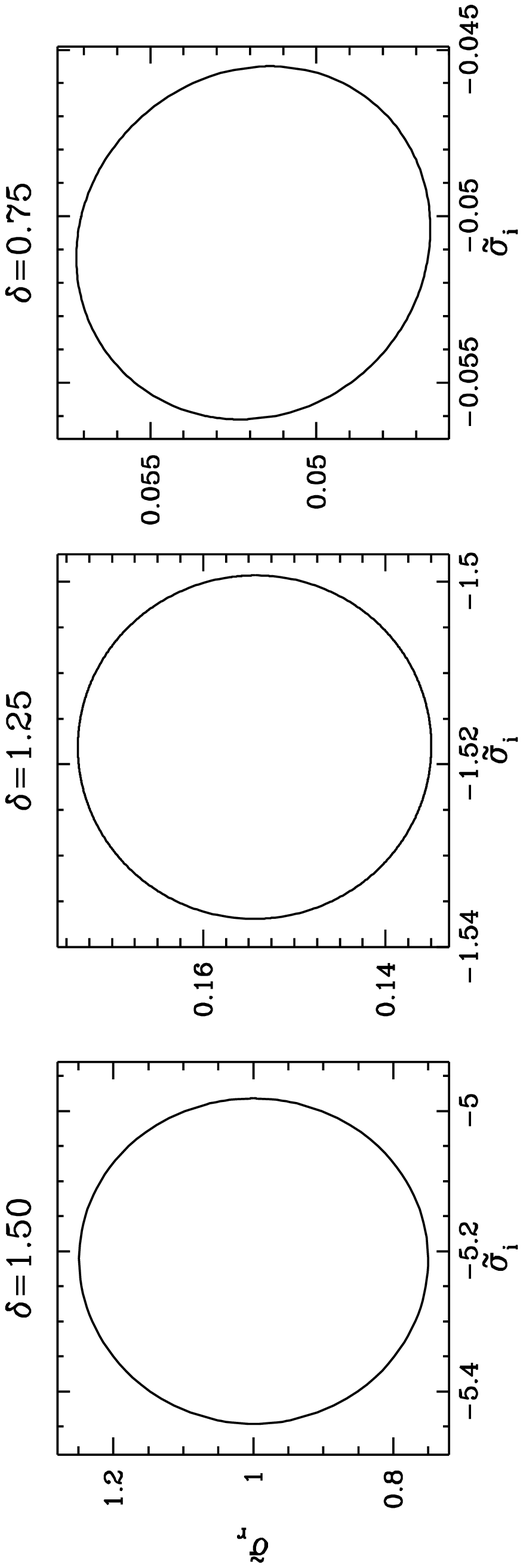}

\clearpage
\plotone{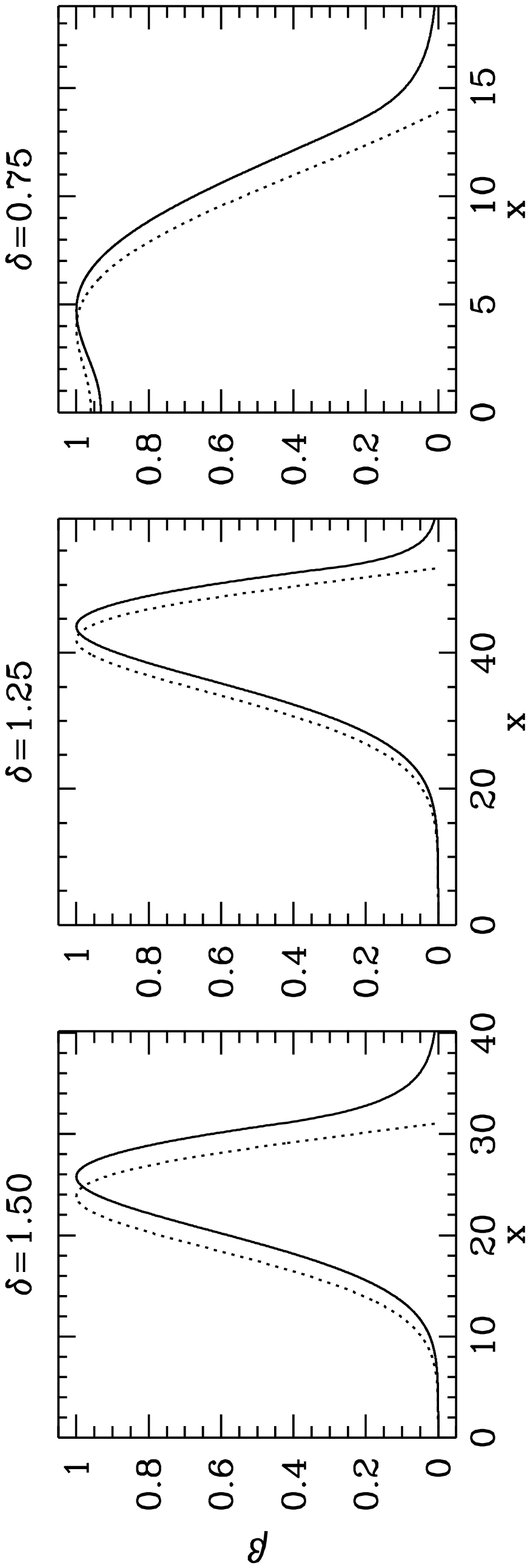}

\clearpage
\plotone{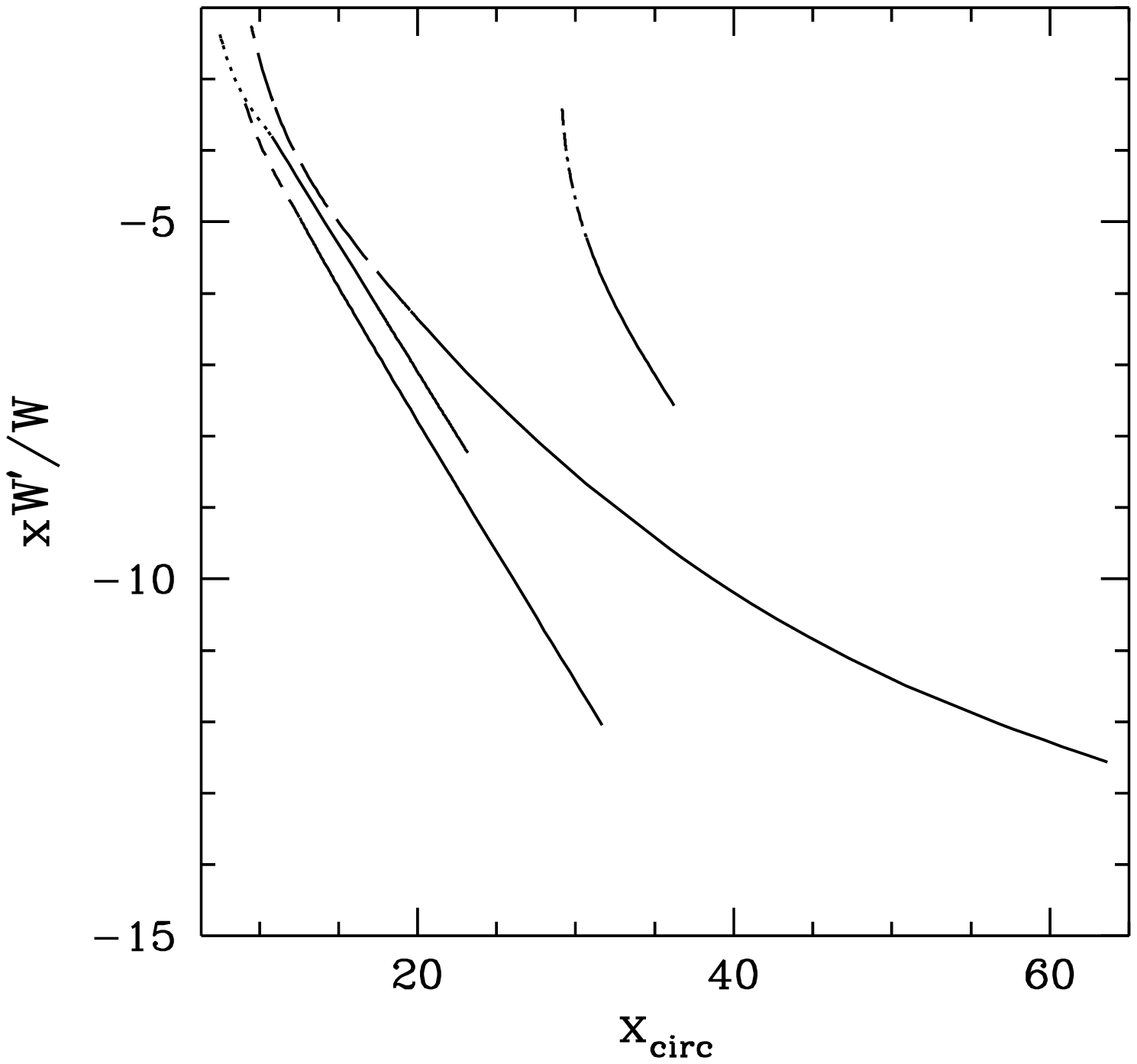}

\clearpage
\plotone{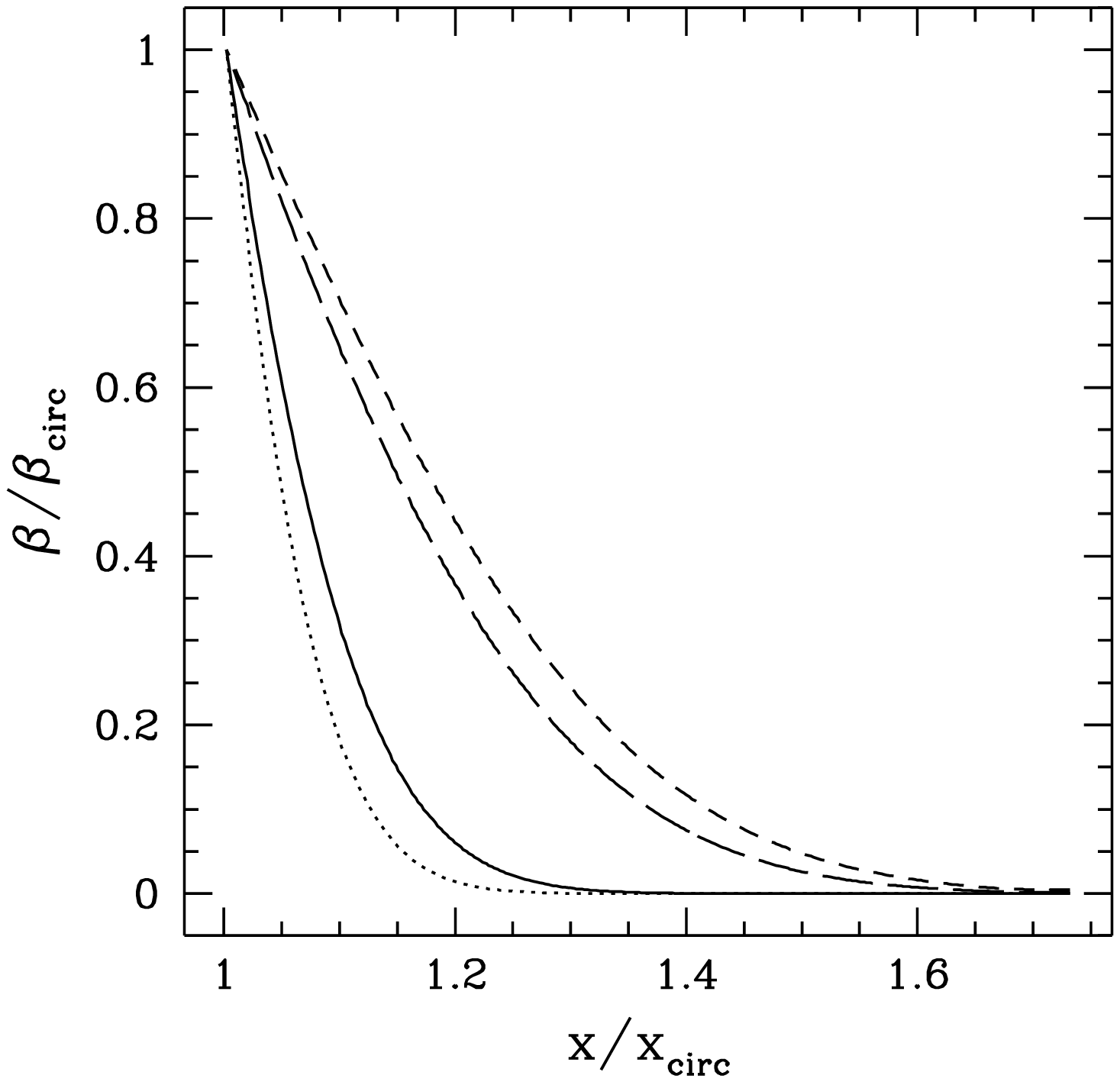}

\clearpage
\plotone{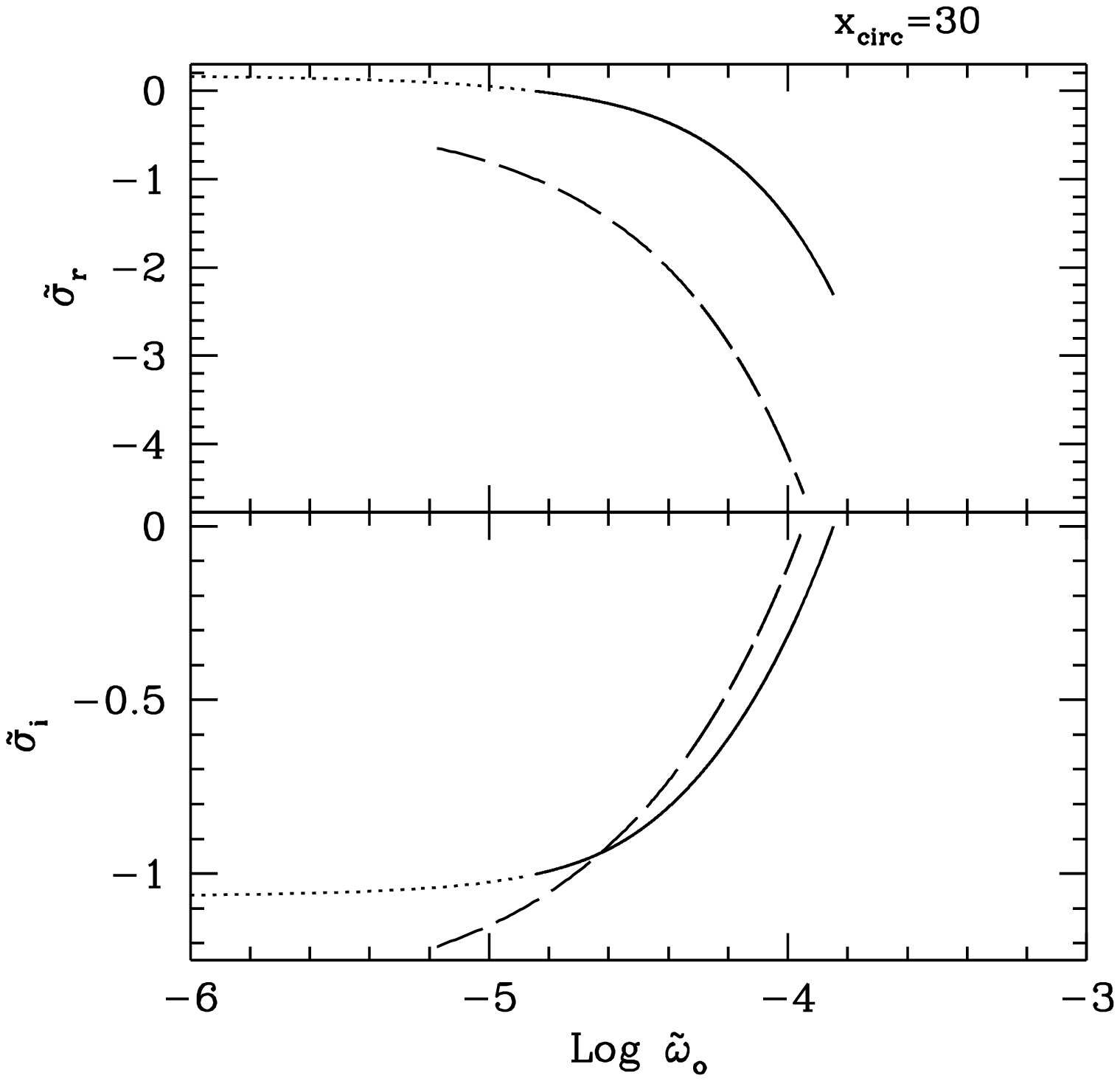}

\end{document}